\documentclass{iopart}

\usepackage[utf8]{inputenc}
\usepackage[english]{babel}
\usepackage{graphicx}
\usepackage{subcaption}
\usepackage[separate-uncertainty=true]{siunitx}
\usepackage{todonotes}
\usepackage{svg}
\usepackage{cite}
\usepackage{hyperref}
\usepackage{array}
\usepackage{caption}
\usepackage{diagbox}
\usepackage{amssymb}

\newcolumntype{L}{>{$}l<{$}}

\begin{document}
\title[]{Vibrational CARS measurements in a near-atmospheric pressure 
plasma jet in nitrogen: \\I. Measurement procedure and results}

\author{J Kuhfeld, N Lepikhin, D Luggenhölscher, U Czarnetzki}

\address{Ruhr University Bochum, Institute for Plasma and Atomic Physics, Germany}
 
\ead{jan.kuhfeld@rub.de}

\begin{abstract}
	The non-equilibrium ro-vibrational distribution functions of molecules in a plasma can 
	heavily influence the discharge operation and the plasma-chemistry. A convenient 
	method for measuring the distribution function is by means of coherent anti-Stokes
	Raman scattering (CARS).
	CARS spectra are measured in a ns-pulsed plasma between two parallel, \SI{1}{mm} 
	spaced molybdenum electrodes in nitrogen at
	\SI{200}{mbar} with pulse durations of \SI{200}{ns} / \SI{250}{ns} and 
	a repetition rate of \SI{1}{kHz}. The CARS spectra are analyzed by a fitting
	routine to extract information about the vibrational excitation of the nitrogen
	molecules in the plasma. It is found that during the discharge the vibrational
	distribution for $v \lesssim 7$ can be described by a vibrational two-temperature distribution 
	function. Additionally, the electric field is measured by the E-FISH method
	during the discharge pulse. It is found to be constant in time after the initial 
	ionization wave with values close to \SI{81}{Td} for the investigated conditions.
	During the afterglow between two discharge pulses a more general 
	fitting approach is used to obtain the population differences of two neighboring
	vibrational states. This allows to capture the more complex vibrational 
	dynamics in that time period. The measurement results are discussed in more 
	detail and compared to simple models in a companion paper \cite{kuhfeld_vibrational_nodate}.
\end{abstract}

% \keywords{coherent anti-Stokes Raman scattering, atmospheric pressure plasma jet, ns-pulsed}

\submitto{\JPD}

\maketitle
% \ioptwocol

\section{Introduction}
\label{sec:intro}
Since several years atmospheric pressure plasmas are a very active research topic with respect to 
multiple practical applications, e.g. plasma assisted ignition\cite{starikovskaia_plasma_2006,adamovich_plasma_2009}, 
plasma catalysis\cite{neyts_understanding_2014,neyts_plasma_2015,whitehead_plasmacatalysis_2016,stewig_excitation_2020} and plasma medicine
\cite{fridman_applied_2008, stoffels_cold_2008,kong_plasma_2009,graves_emerging_2012}.
Many of these applications depend on the unique chemical properties of atmospheric pressure
plasmas which have their roots in the different non-equilibrium conditions in these 
kinds of discharges\cite{fridman_plasma_2008}. 
A very important non-equilibrium can be found in the vibrational distribution functions (VDFs) 
of molecular plasmas. Strong excitation through resonant interaction with the plasma electrons 
and the anharmonicity of the vibrational potential usually lead to a strong overpopulation 
of the higher vibrationally excited states compared to a Boltzmann distribution\cite{fridman_plasma_2008}. 
On the other hand the translational and rotational temperatures can be very low 
(close to room temperature) due to the low coupling of the electrons to the translational 
and rotational modes of the molecules and slow vibrational-translational (V-T) relaxation 
\cite{fridman_plasma_2008}.
This means that energy to be used in chemical reactions can be stored in the gas 
without increasing the translational temperature. 
There are several ways how this can be leveraged in industrial applications. 
For example catalysts could potentially be used at gas temperatures 
below their traditional operating temperature which could increase their durability 
and efficiency\cite{urbanietz_non-equilibrium_2018,stewig_excitation_2020}.
Another process of interest over the last years is carbon dioxide dissociation by step wise 
vibrational excitation which might be more efficient than direct dissociation
by electron impact\cite{aerts_influence_2012,aerts_carbon_2015,bogaerts_plasma-based_2015}. \\
All these examples clearly state an interest in tailoring the vibrational excitation 
in a plasma for the specific application by using different plasma sources, 
e.g. (surface) dielectric barrier discharges, plasma jets or micro-structured array devices. 
For this purpose the influence of the plasma parameters on the VDF needs to be understood in detail, 
so kinetic modelling as well as detailed measurements need to be 
performed in parallel for different discharge types and conditions. \\
Measurements of the VDF can be performed for some gases or vibrational modes 
by TDLAS or FTIR \cite{urbanietz_non-equilibrium_2018, klarenaar_time_2017,stewig_excitation_2020}, 
but this is not possible in other 
cases as e.g. nitrogen or the symmetric stretch mode of carbon dioxide. 
In these cases other methods need to be employed like spontaneous Raman scattering or 
coherent anti-Stokes Raman scattering (CARS). A detailed description of previously performed CARS and
spontaneous Raman scattering measurements in plasmas is given in \cite{lempert_coherent_2014}
so here only a short summary is provided. Some of the first CARS measurements in
nitrogen plasmas known to the authors were performed in low pressure DC discharges 
for example by Shaub \etal \cite{shaub_direct_1977} who also developed a scheme
for the evaluation of CARS spectra with non-Boltzmann vibrational distributions.
Their scheme is valid under the assumption that the bulk of the molecules follows
a Boltzmann distribution and only a very small number is excited to higher vibrational
states, leaving the distribution of the bulk. This might be the case for low 
pressure DC discharges but for pulsed high current plasmas a significant amount, 
i.e. in the range of some \SI{10}{\percent}, 
of the molecules can be excited during one pulse.\\
Measurements in pulsed nitrogen plasmas where performed among others by 
Valyansky \etal\cite{valyanskii_studies_1984}, Deviatov \etal 
\cite{deviatov_investigation_1986}, Vereshchagin \etal \cite{vereshchagin_cars_1997},
and Montello \etal \cite{montello_picosecond_2013}. 
In general the vibrational excitation in these works can be described on three 
time scales: The excitation during the discharge (mainly) by resonant electron
collisions usually happens on time scales of nanoseconds, 
the redistribution of vibrational-vibrational (V-V) transfer collisions in the first microseconds after the
pulse and the loss of vibrational excitation by diffusion or 
deactivation at the walls, typically some 
hundreds of microseconds or a few milliseconds after the discharge. 
Depending on the experimental conditions, loss through V-T collisions can also 
be an important mechanism. In some of the mentioned works an increase of the lower
lying vibrational excited states in the first microseconds of the afterglow was observed which could not be explained purely
by V-V transfer mechanisms leading the authors to the conclusion that quenching 
from metastable electronic states \cite{deviatov_investigation_1986,montello_picosecond_2013}
might have a significant impact on the vibrational excitation. This was not observed
in \cite{valyanskii_studies_1984} and \cite{vereshchagin_cars_1997} where 
no significant deviation from the populations predicted by V-V models was seen.
More recently Yang \etal \cite{yang_simulation_2016} tried to explain the measurement results by numerical
modelling but were also not able to reproduce the increase of the total number
of vibrational quanta observed in the measurements. Consequently, the mechanism
behind this process is still an open research question and will certainly be different
for different discharge conditions and geometries.\\
In this work measurements of the VDF of nitrogen are performed in an near atmospheric 
pressure ns-pulsed plasma jet with plan parallel molybdenum electrodes  
similar to the ones used in \cite{schregel_ignition_2016}.
In contrast to some earlier works mentioned above  
the vibrational distribution is not determined by comparing the intensity of the different 
peaks in the CARS spectrum. Instead, it is determined by fitting a calculated spectrum 
like it is traditionally done for rotational temperature measurements by CARS \cite{farrow_comparison_1982},
but on a larger spectral range to capture multiple vibrational transitions.

\section{CARS method}
\label{sec:CARS}
\subsection{CARS spectra and fitting routine}
\label{sec:CARS_spectra}
\begin{figure}
	\centering
	\includegraphics[width=\linewidth]{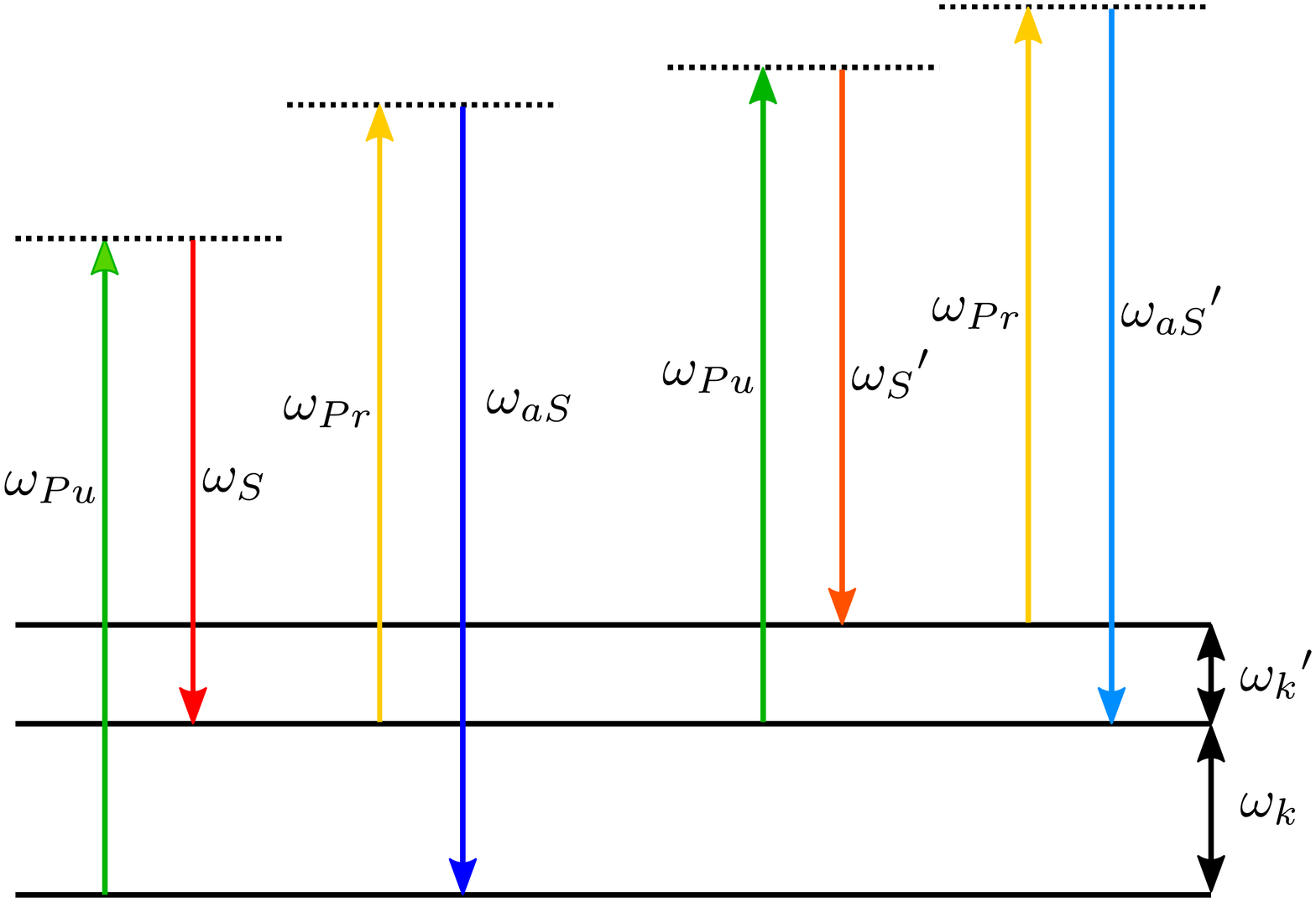}
	\caption{Energy scheme of the CARS process. $\omega_{pu}$, $\omega_S$,
	$\omega_{pr}$ and $\omega_{aS}$ denote the angular frequencies of the 
	pump, Stokes, probe and anti-Stokes beams respectively. ${\omega_k}$ is the
	angular frequency corresponding to a vibrational transition of the probed molecule.
	The prime denotes a second transition with lower energy (e.g. because of the anharmonicity
	of the vibrational potential) enforcing a slightly different Stokes ${\omega_S}'$ and 
	anti-Stokes ${\omega_{aS}}'$ frequency.}
	\label{fig:energy_scheme}
\end{figure}
There are multiple reviews about coherent anti-Stokes Raman scattering (CARS) as 
a diagnostic method for temperature and concentration measurements in gases and other media
\cite{druet_cars_1981,eesley_coherent_1979,el-diasty_coherent_2011, marowsky_coherent_1992,szymanski_raman_1970},
so here we will only give a short summary of the aspects relevant for this work.
Coherent anti-Stokes Raman scattering is a non-linear optical process of 
third order, in which three electromagnetic waves with angular frequencies 
$\omega_{Pu}$, $\omega_S$ and $\omega_{Pr}$ induce a non-linear polarization in 
a medium at frequency $\omega_{aS} = \omega_{Pu} + \omega_{Pr} - \omega_S$.
$\omega_{Pu}$, $\omega_S$, $\omega_{Pr}$ and $\omega_{aS}$ are called pump, Stokes, probe
and anti-Stokes beam respectively. The energy scheme of the process is illustrated in 
figure \ref{fig:energy_scheme} for two transitions. The process is resonant if $\omega_{Pu} - \omega_S$ is 
close to a state transition of the used medium. In this work the transition energy 
$\hbar\omega_k$ belongs to vibrational transitions of nitrogen.
The intensity of the generated anti-Stokes beam is proportional to the square of the 
so called CARS susceptibility and the product of the three input laser intensities: 
\begin{equation}
	I_{aS}(\omega_R) \propto |\chi_{CARS}(\omega_R)|^2 I_{Pu} I_{Pr} I_S(\omega_R)
	\label{eq:as_intensity}
\end{equation}
where $\omega_R = \omega_{Pu} - \omega_S$ is the Raman shift. 
The CARS susceptibility in the pressure broadening regime for parallel polarized
light far from electronic resonances can be written as
\begin{equation}
	\chi_{CARS}(\omega) = \chi_{NR} + \chi_R(\omega) = \chi_{NR} + \sum_k \frac{a_k}{\omega_k - \omega - i\Gamma_k}
	\label{eq:chi_cars}
\end{equation}
where the sum includes all possible ro-vibrational transitions $k$ with the energy difference
$\hbar\omega_k=E(v_{k,f}, J_{k,f}) - E(v_{k,i}, J_{k,i})$ and the linewidth (HWHM) $\Gamma_k$.
The ro-vibrational energies are calculated with the Dunham coefficients given in table \ref{tab:dunham} by
\begin{equation}
	E(v,J) = \sum_{i,j} Y_{ij} (v+\frac{1}{2})^i (J(J+1))^j
\end{equation}

\begin{table*}
	\centering
	\caption{Dunham coefficients $Y_{ij}$ in units of \si{cm^{-1}}. The data marked with 
	$^{(a)}$ is from \cite{lavorel_dunham_1988}, $^{(b)}$ from \cite{gilson_redetermination_1980}
	and $^{(c)}$ from \cite{huber_molecular_1979,luthe_algorithms_1986}\label{tab:dunham}}
	\begin{tabular}{l||cccc}
		\diagbox{i}{j}	&	0								&	1								&	2						&	3						\\ \hline\hline	
		0	&	0								&	1.998 $^{(c)}$							&	$-5.76\times 10^{-6}$ $^{(c)}$	&	0						\\
		1	&	2358.535 $^{(a)}$				&	$-0.017249$ $^{(a)}$			&	$-8.32\times 10^{-9}$ $^{(b)}$	&	$-2.58\times 10^{-13}$ $^{(b)}$	\\
		2	&	-14.3074 $^{(a)}$				&	$-3.24\times 10^{-5}$ $^{(a)}$	&	$-4.2\times 10^{-10}$ $^{(b)}$	&	0						\\
		3	&	$-4.98\times 10^{-3}$ $^{(a)}$	&	0								&	0						&	0					\\	
		4	&	$-1.22\times 10^{-4}$ $^{(a)}$ 	&	0								&	0						&	0						
	\end{tabular}
\end{table*}

With the assumptions above the amplitude of the lines depend on the population difference
between the lower and the upper state of the transition $\Delta N_k = N_{k,l} - N_{k,u}$
and the differential cross section for spontaneous Raman scattering:
\begin{equation}
	a_k = \left( \frac{c^4}{\hbar \omega_S^4} \right) \Delta N_k \left . \frac{\textrm{d}\sigma}{\textrm{d}\Omega} \right|_k.
	\label{eq:chi_ampl}
\end{equation}
The Raman cross section for the Q branch ($\Delta v = 1, \Delta J =0$) can be written as\cite{luthe_algorithms_1986, farrow_comparison_1982} 
\begin{equation}
	\left.\frac{\textrm{d}\sigma}{\textrm{d}\Omega}\right|_Q = 
	\left(\frac{\omega_S}{c}\right)^4\frac{\hbar}{2\omega_0}
	\left[\frac{\alpha'^2}{M} + \frac{4}{45} \frac{\gamma'^2}{M}b^J_J\right](v+1)
	\label{eq:cross_section_Q}
\end{equation}

\begin{figure}
	\centering
	\includegraphics[width=\linewidth]{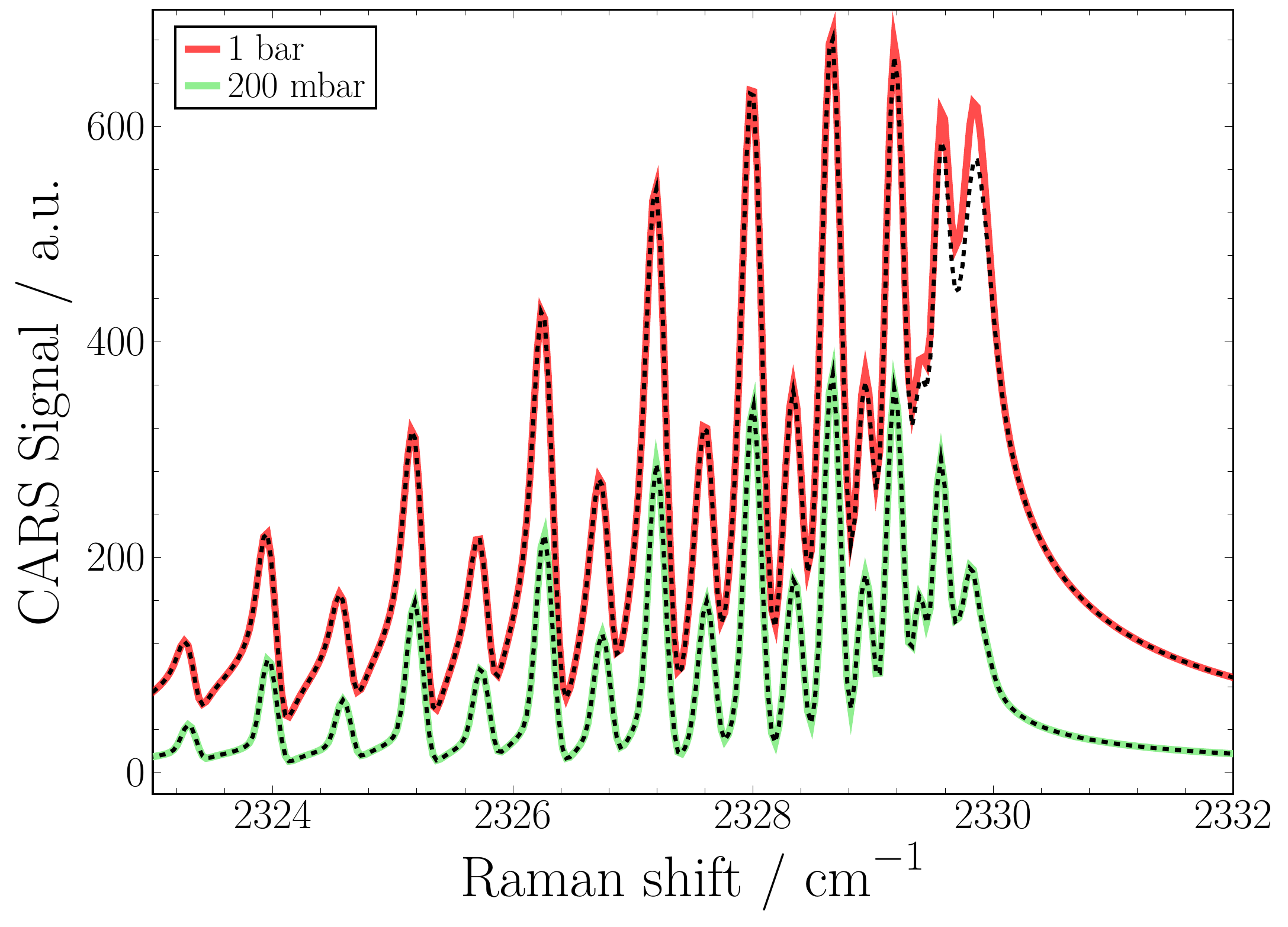}
	\caption{Comparison of spectra calculated with the isolated lines model (dashed)
	compared to calculations with the exponential gap model (solid) in pure 
	nitrogen ($T=\SI{350}{K}$) at \SI{1}{bar} (red) and \SI{200}{mbar} (blue).}
	\label{fig:iso_vs_exp}
\end{figure}

The cross sections for the O and S branch are given by\cite{luthe_algorithms_1986,farrow_comparison_1982}
\begin{equation}
	\left.\frac{\textrm{d}\sigma}{\textrm{d}\Omega}\right|_O = 
	\left(\frac{\omega_S}{c}\right)^4\frac{\hbar}{2\omega_0}
	\frac{4}{45} \frac{\gamma'^2}{M}b^J_{J+2}(v+1)C_O(J)
	\label{eq:cross_section_O}
\end{equation}
\begin{equation}
	\left.\frac{\textrm{d}\sigma}{\textrm{d}\Omega}\right|_S = 
	\left(\frac{\omega_S}{c}\right)^4\frac{\hbar}{2\omega_0}
	\frac{4}{45} \frac{\gamma'^2}{M}b^J_{J-2}(v+1)C_S(J)
	\label{eq:cross_section_S}
\end{equation}
with the centrifugal force corrections\cite{farrow_comparison_1982}
\begin{equation}
	C_O(J) = \left(1+4\frac{B_e}{\omega_e} \mu (2J-1)\right)^2
\end{equation}
\begin{equation}
	C_S(J) = \left(1-4\frac{B_e}{\omega_e} \mu (2J+3)\right)^2
\end{equation}
In the equations above $\omega_0$ is the oscillator frequency of the 
molecule, $\frac{\alpha'^2}{M}$ and $\frac{\gamma'^2}{M}$ are the squared isotropic and anisotropic
derived polarizabilities over the reduced mass of the vibration, $B_e$ and $\omega_e$ the Herzberg molecular parameters
and 
\begin{equation}
	b_{J}^J = \frac{J(J+1)}{(2J-1)(2J+3)}
\end{equation}
\begin{equation}
	b_{J+2}^{J} = \frac{3(J+1)(J+2)}{2(2J+1)(2J+3)}
\end{equation}
\begin{equation}
	b_{J-2}^{J} = \frac{3J(J-1)}{2(2J+1)(2J-1)}
\end{equation}
are the Placzek-Teller coefficients for diatomic molecules\cite{luthe_algorithms_1986}.\\
The constants $\frac{\alpha'^2}{M}$ and $\frac{\gamma'^2}{M}$ in \eref{eq:cross_section_Q},\eref{eq:cross_section_O} and \eref{eq:cross_section_S}
are calculated following the approach of \cite{farrow_comparison_1982} from 
experimental measurements of the Q branch cross section for $v=0\rightarrow 1$

\begin{equation}
		\left.\frac{\mathrm{d} \sigma}{\mathrm{\Omega}}\right|_{Q,v=0} \approx 
		\frac{\hbar \omega_S^4}{2\omega_0c^4}\frac{\alpha'^2}{M}
\label{eq:alpha}
\end{equation}
and the Q branch depolarization ratio
\begin{equation}
	\rho_J = \frac{3b_{JJ} (\gamma'/\alpha')^2}{45 + 4b_{JJ} (\gamma'/\alpha')^2}.
\end{equation}
While the line positions $\omega_k$ in \eref{eq:chi_cars} depend only on the molecular
parameters, the linewidths $\Gamma_k$ depend on the gas mixture, the pressure and the temperature.
There exist multiple scaling laws for the linewidths in nitrogen, for example the so called polynomial-differential
exponential gap law (PDEGL) or the modified exponential gap law (MEGL)\cite{clark_advances_1988}.
In this work the linewidths are calculated with\cite{clark_advances_1988}
\begin{equation}
	\Gamma_j = \sum_{i>j}\gamma_{ij}.
	\label{eq:Gamma_j}
\end{equation}
Note the different index notation compared to the one used in \eref{eq:chi_cars} where the 
index $k$ corresponds to a transition between to ro-vibrational states. 
In \eref{eq:Gamma_j} $i,j$ correspond to rotational quantum numbers. The relation between
the two notations is that - for a \mbox{Q branch} transition - $j$ is the rotational quantum 
number which stays constant during the transition $k$.
For O and S branch transition the rotational quantum numbers $u$ and $l$ for the 
upper and lower state respectively are different and the line width for 
transition $k$ is calculated as 
\begin{equation}
	\Gamma_k = \frac{1}{2} \left( \Gamma_u + \Gamma_l\right).
\end{equation}
$\gamma_{ij}$ in the above equations is the collisional transfer rate for rotational states $j\rightarrow i$.
If the rotational states are Boltzmann distributed detailed balance requires for
the back and forth rates:
\begin{equation}
g(i)\exp(-E_i/k_BT)\gamma_{ji} = g(j)\exp(-E_j/k_BT)\gamma_{ij}	
\end{equation}
where $g$ are the statistical weights for the rotational states. 
In this work the MEGL is used which gives for the uprates\cite{clark_advances_1988}
\begin{equation}
	\hspace*{-2cm}
	\gamma_{ji}=p\alpha\frac{1-e^{-m}}{1-e^{-mT/T_0}}\left(\frac{T_0}{T}\right)^{1/2}
	\left(\frac{1+1.5E_i/k_BT\Delta}{1+1.5E_i/k_BT}\right)^2
	\exp(-\beta|\Delta E_{ji}|/k_BT)
\end{equation}
where $p$ is the pressure in bar and $T$ the gas temperature (in the following discussions assumed to be equal to the rotational 
temperature). The parameters for nitrogen are
$\alpha=\SI{0.0231}{\per\centi\metre\per\bar}$, $\beta=1.67$, $\Delta=1.21$, $m=0.1487$ and $T_0=\SI{295}{\kelvin}$.\\
In this context it should be mentioned that the line profile in  \eref{eq:chi_cars}
is only valid in low or medium pressures as it assumes that the lines are isolated.
For pressures close to atmospheric pressure this is not necessarily true and more complex
line shapes are required like they are given in the rotational diffusion model
or the exponential gap model\cite{clark_advances_1988}. To verify if the isolated line model is appropriate 
under the conditions in this work (nitrogen at \SI{200}{mbar}) nitrogen CARS spectra of the
Q branch transition $v=0\rightarrow 1$
are calculated for \SI{200}{mbar} and \SI{1}{bar} with the CARSFT \cite{palmer_carsft_1989-1} software for both 
the isolated lines and the more complex exponential gap model. The results are compared
in figure \ref{fig:iso_vs_exp}. While there is a noticeable difference between the
isolated lines and the exponential gap model for \SI{1}{bar} at low rotational quantum numbers,
the curves at \SI{200}{mbar} are indistinguishable. As conclusion the isolated lines model 
is sufficient for the conditions investigated here.

As seen in \eref{eq:chi_cars} and \eref{eq:chi_ampl} the intensity of the 
generated anti-Stokes beam depends on the population density differences of the participating
states. This means by scanning the Stokes beam over multiple
transitions and measuring the anti-Stokes intensity, information about the 
ro-vibrational distribution functions can be extracted. Equivalently, a Stokes
beam with a continuous broadband spectrum can be used to record the whole anti-Stokes
spectrum with a single laser shot as is done in this work.
One way of gaining information is by fitting theoretically
calculated spectra to the measured ones, which of course depends
on correct models and precise data used in the calculation.
This approach is used in rotational temperature measurements 
where the rotational structure needs to be resolved and lines can be overlapping at certain pressures or 
even motional narrowing \cite{hall_pressure-induced_1980} can occur. \\
For most plasma sources the vibrational and rotational temperature are
typically not the same. In these cases spectrometers with a larger spectral range are
used to monitor multiple vibrational lines. This usually results in a lower
spectral resolution so that the rotational structure of the lines can not be
resolved. Then the accuracy of the rotational temperature is naturally lower. \\
A popular approach to determine the vibrational distribution from these low-resolution
spectra is to first normalize the spectrum to either the Stokes laser spectrum or
a non-resonant spectrum from another gas without vibrational resonances in that 
wavelength region. Then the square root of the normalized spectrum is taken and 
for each transition the integral over the corresponding peak $I_{v, v+1}$ is calculated.
Under the assumption that the non-resonant and the real part of the resonant 
susceptibility are negligible compared to the imaginary part of the resonant 
susceptibility, $\chi_{NR},\text{Re} (\chi_R) \ll \text{Im} (\chi_R) $, it follows that  
\begin{equation}
	I_{v, v+1} \propto \Delta N_{v,v+1}(v+1),
\end{equation}
where the factor $v+1$ accounts for the dependence of the Raman cross section on the vibrational
quantum number (see \eref{eq:cross_section_Q}). \\
As this approach relies heavily on the made assumptions, which are certainly not valid for small
admixtures of the probed gas species \cite{clark_advances_1988} and probably also not true for the weak signals of the higher vibrational states,
in this work the population densities are extracted by fitting the full theoretical spectra to the measured ones. \\
To calculate the population differences in the susceptibility expression 
different models are used depending on the corresponding conditions.
One possibility would be to assume that the vibrational states are Boltzmann distributed
but with a different temperature than the rotational states like it was done for 
example by Messina \etal\cite{messina_study_2007}:
\begin{equation}
	N(v,J) = \frac{g(J)}{Z} 
	\mathrm{e}^{-\frac{E(v,J)-E(v,0)}{k_BT_{rot}}} 
	\times \mathrm{e}^{-\frac{E(v,0)-E(0,0)}{k_BT_{vib}}}
	\label{eq:N_boltz}
\end{equation}
with the partition function
\begin{equation}
	Z = \sum_{v,J} g(J)
	\mathrm{e}^{-\frac{E(v,J)-E(v,0)}{k_BT_{rot}}} 
	\times \mathrm{e}^{-\frac{E(v,0)-E(0,0)}{k_BT_{vib}}}
	\label{eq:Z_boltz}
\end{equation}
and $T_{vib} \neq T_{rot}$. Another approach would be the Treanor distribution
assuming dominant V-V transfer collisions between the molecules:
\begin{equation}
	N(v,J) = \frac{g(J)}{Z} 
	\mathrm{e}^{-\frac{E(v,J)-E(v,0)}{k_BT_{rot}}} 
	\times \mathrm{e}^{-\frac{v(E(1,0)-E(0,0))}{k_BT_{vib}}}
	\times \mathrm{e}^{-\frac{v(E(v,0)-E(1,0))}{k_BT_{rot}}}
	\label{eq:N_treanor}
\end{equation}
with
\begin{equation}
	Z = \sum_{v,J} g(J)
	\mathrm{e}^{-\frac{E(v,J)-E(v,0)}{k_BT_{rot}}} 
	\times \mathrm{e}^{-\frac{v(E(1,0)-E(0,0))}{k_BT_{vib}}}
	\times \mathrm{e}^{-\frac{v(E(v,0)-E(1,0))}{k_BT_{rot}}}.
	\label{eq:Z_treanor}
\end{equation}
The degeneracy of the rotational states is given by 
\begin{equation}
	%g(J) = \cases{6(2J+1), \text{for } J \text{ even}}{3(2J+1),  \text{for } J \text{ odd}}
	g(J) = \left\{\begin{array}{@{\kern2.5pt}lL}
			\hfill 6(2J+1),& \text{if } J \text{ even} \\
			\hfill 3(2J+1),&  \text{if } J \text{ odd.}
		\end{array}\right.
\end{equation}
Both distributions assume a single vibrational temperature and 
are for the temperatures and relatively low vibrational 
states measured in this work essentially indistinguishable
considering the uncertainty of the CARS measurements. \\
As in previous measurements of the ro-vibrational distribution \cite{montello_picosecond_2013,vereshchagin_cars_1997,valyanskii_studies_1984,deviatov_investigation_1986},
here, it was found that a single vibrational temperature is not sufficient to fully
describe the vibrational distribution for $v>3$. Therefore, here a distribution 
function considering two vibrational temperatures, a cold and a hot one, is introduced 
similar to the one used in \cite{du_emission_2017}. The subtle difference compared 
to \cite{du_emission_2017} is that the hot part of the distribution only includes 
states with $v\geq 1$. This is rooted in the interpretation that the hot distribution 
is made up by molecules which are excited during the current discharge cycle.
The further motivation of this two-temperature distribution function (TTDF) is discussed in more detail 
in the companion paper \cite{kuhfeld_vibrational_nodate} and here only the formula is given:
\begin{eqnarray}
	N(v,J; T_{r}, T_{vib,c}, T_{vib,h}, R_{h}) = 
	g(J) \mathrm{e}^{-\frac{E(v,J) - E(v,0)}{k_BT_{rot}}} \nonumber \\
	\times\Bigl[
		\frac{1-R_h}{Z_c}\mathrm{e}^{-\frac{E(v,0)-E(0,0)}{k_BT_{vib,c}}} 
		+ \underbrace{\frac{R_h}{Z_h}\mathrm{e}^{-\frac{E(v,0)-E(0,0)}{k_BT_{vib,h}}}}_{\text{for }v>0}
	\Bigr]
	\label{eq:N_twotemp}
\end{eqnarray}
with the partition function for the cold molecules
\begin{equation}
	Z_c = \sum_{v,J} g(J) \mathrm{e}^{-\frac{E(v,J) - E(v,0)}{k_BT_{rot}}}
	\times\mathrm{e}^{-\frac{E(v,0)-E(0,0)}{k_BT_{vib,c}}}
	\label{eq:Z_c}
\end{equation}
and
\begin{equation}
	Z_h = \sum_{v>0,J} g(J) \mathrm{e}^{-\frac{E(v,J) - E(v,0)}{k_BT_{rot}}}
	\times\mathrm{e}^{-\frac{E(v,0)-E(0,0)}{k_BT_{vib,h}}}
	\label{eq:Z_h}
\end{equation}
for the hot molecules.\\
The last condition investigated in this work is several microseconds after the discharge pulse. 
On this time scales the V-V collisions determine the temporal behavior and 
the two-temperature description is not valid anymore. 
Indeed, all of the distributions mentioned above fail in this regime.
To make a distribution function independent analysis possible only the Q branch, $\Delta v=1, \Delta J=0$ is considered.

%afterglow
If the rotational states in a vibrational level $v$ follow a Boltzmann distribution 
$f_v(J, T_{v,rot})$ with rotational temperature $T_{v,rot}$, 
the population difference for a given transition $v\rightarrow v+1; J$ is
\begin{eqnarray}
	\Delta N_{v,v+1;J} 	&= N_v f_v(J, T_{v,rot}) - N_{v+1} f_{v+1}(J,T_{v+1,rot}) \\\nonumber
						&\approx (N_v - N_{v+1}) f(J, T_{rot}) \\\nonumber
						&= \Delta N_{v, v+1} f(J, T_{rot}) 
\end{eqnarray}
Here, in the second step it is assumed that the rotational temperature 
in the different vibrational states is the same and that the vibrational corrections 
to the rotational distributions are negligible. In this way the fitting parameters 
concerning the vibrational states are reduced to $\Delta N_{v, v+1}$. 

\subsection{Benchmark}
\label{sec:benchmark}
\begin{figure}
	\centering
		\includegraphics[width=\linewidth]{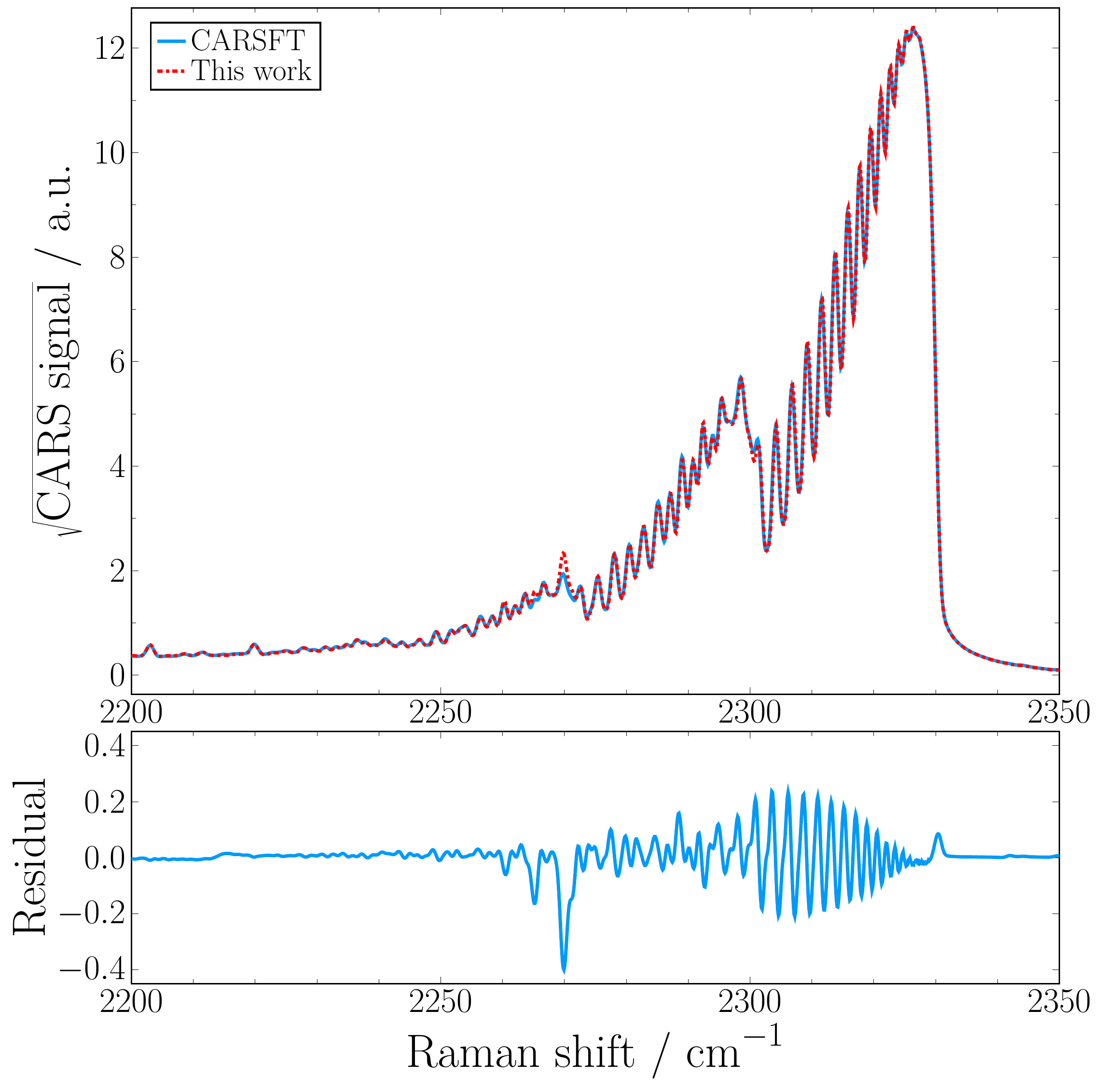}
		\caption{CARSFT \cite{palmer_carsft_1989-1} spectrum at $T=\SI{2000}{K}$ and fit by the code developed in this work with independent 
		vibrational and rotational temperatures $T_{vib}= \SI{2000}{K}$ and $T_{rot}= \SI{1996}{K}$.\label{fig:CARSFT_fit}}
\end{figure}
\begin{figure}
	\centering
		\includegraphics[width=\linewidth]{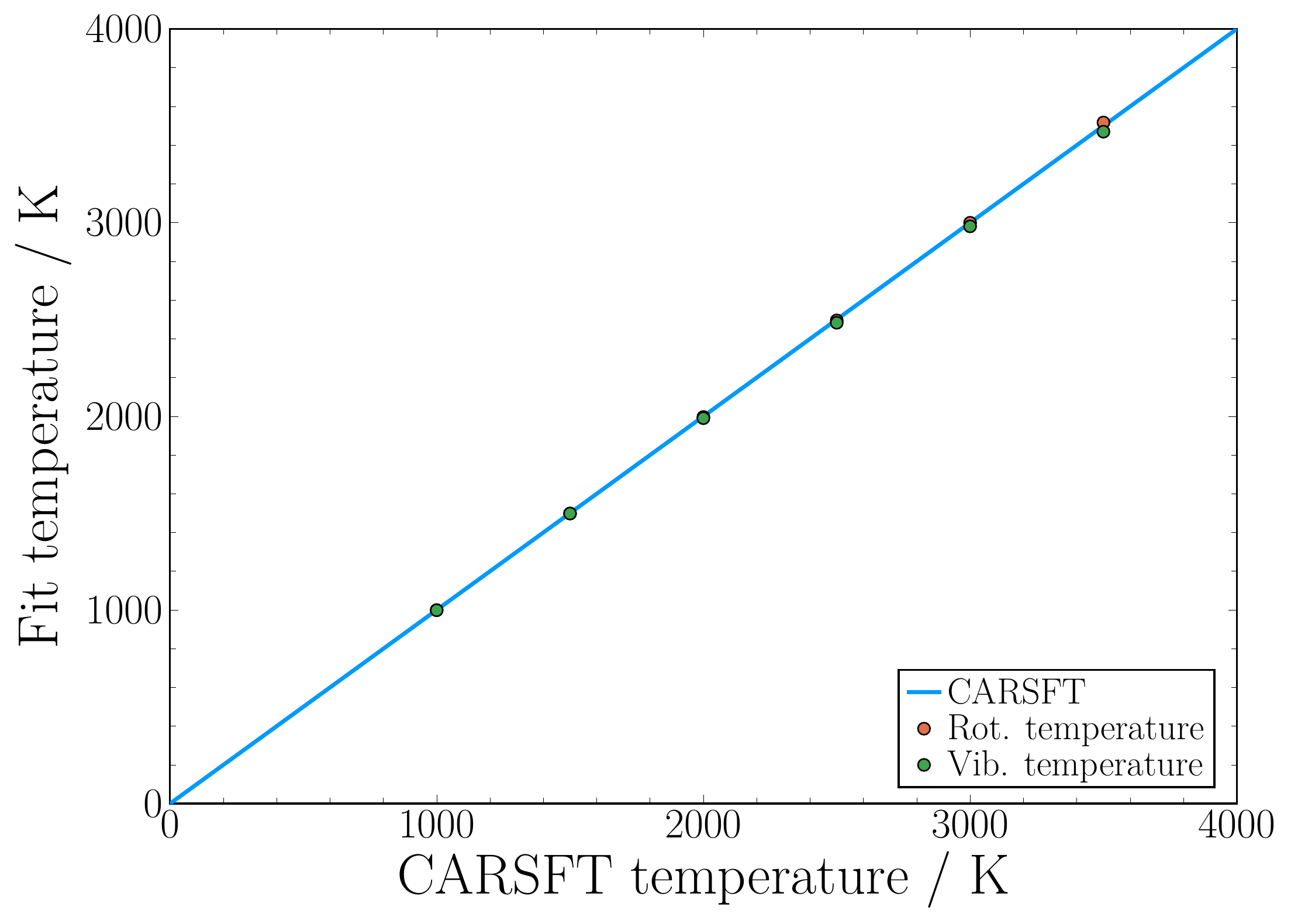}
		\caption{Comparison of the vibrational and rotational temperatures extracted by the fitting routine and the 
		equilibrium temperatures used to produce the CARSFT spectra. The diagonal line corresponds to perfect agreement 
		between theoretical and fitted temperatures.\label{fig:CARSFT_comp}}
\end{figure}
To benchmark our CARS code and the determination of vibrational temperatures from the fitting parameters 
several spectra are generated by the popular CARSFT code \cite{palmer_carsft_1989-1}
with a spectral resolution similar to our experiments 
for thermal equilibrium conditions and evaluated by our fitting routines assuming Boltzmann distributed vibrational 
and rotational states where $T_{rot}$ and $T_{vib}$ are 
independent of each other. An example fit for an equilibrium temperature of 
\SI{2000}{K} is shown in figure \ref{fig:CARSFT_fit}. 
In figure \ref{fig:CARSFT_comp} the fitting results are shown for 
several spectra generated by the CARSFT code. 
The agreement for both the rotational and the vibrational temperature is very good over the whole range 
,and it is reasonable to assume that the code used in this work accurately describes real CARS spectra in the 
given temperature range even if $T_{rot}\neq T_{vib}$.

\section{Electric field measurements}
To understand the plasma processes leading to the measured ro-vibrational distributions
knowledge about the electric field in the discharge is essential.
To measure the electric field value in the discharge the Electric Field Induced 
Second Harmonic generation (E-FISH) technique 
\cite{simeni_electric_2018, chng_electric_2019,chng_electric_2019-1, orr_measurements_2020,adamovich_nanosecond_2020, huang_surface_2020,simeni_simeni_electric_2018, goldberg_electric_2018, dogariu_species-independent_2017, lepikhin_electric_2020}
is used. 
It is a third-order nonlinear process  involving  an electric field of a light 
wave with frequency $ \omega $ (laser emission) and an external applied electric 
field, $ E $. As a result, light with doubled frequency is generated 
with intensity proportional to $ E^2 $:
\begin{equation}
I^{(2\omega)}=kN^2I^2_{laser}E^2, 
\label{eq_I_EFISH}
\end{equation}
\noindent
where $ k $ is a constant value depending on the gas mixture, its susceptibility and the optical system, 
$ N $ is the gas density and $ I_{laser} $ the laser pulse energy.

\begin{figure}[h!]
	\centering{\includegraphics[width=\linewidth]{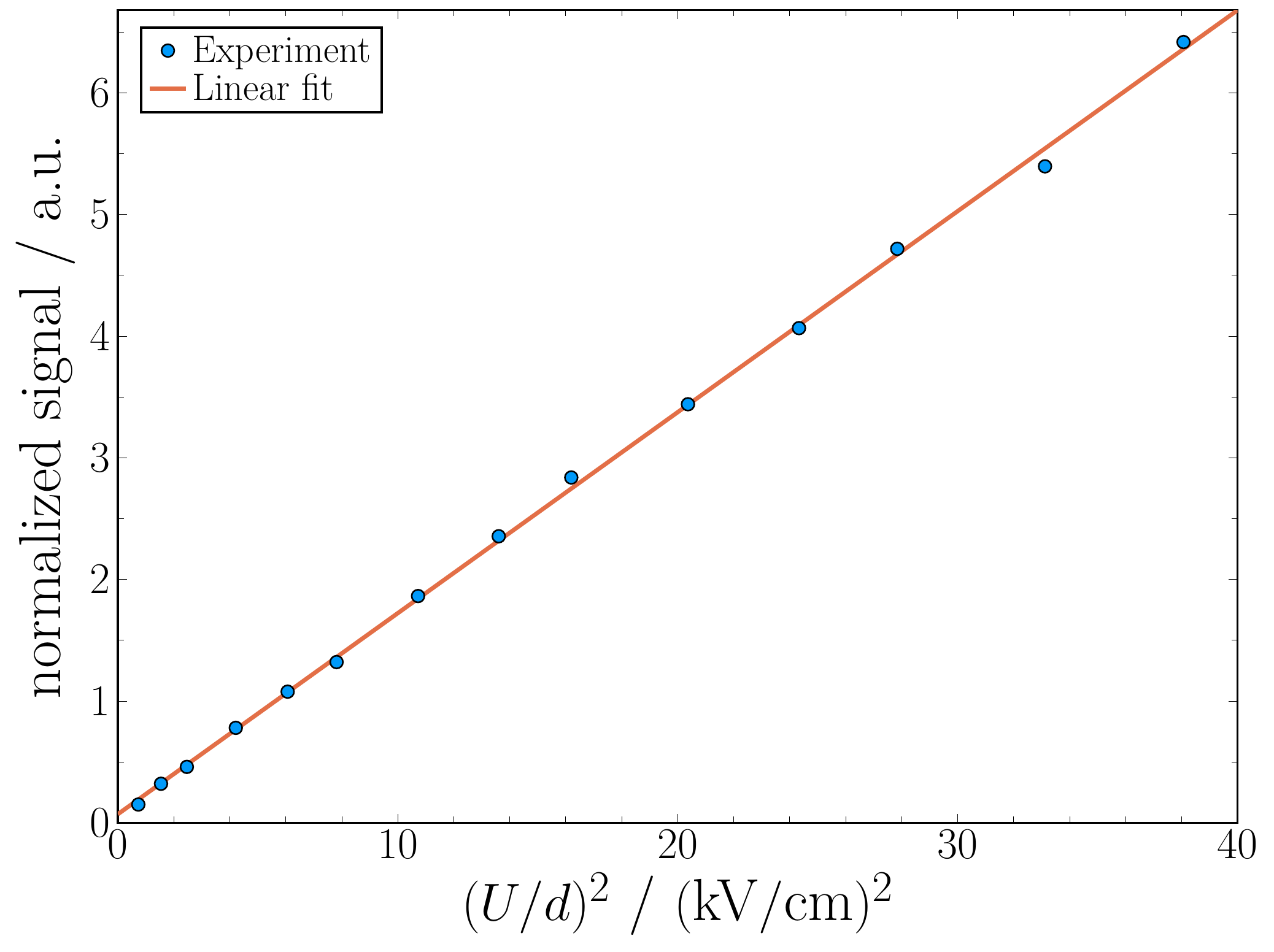}}
	\caption{Calibration curve used for E-FISH measurements obtained 
	with 150~ns rectangular pulses at 10 Hz: measured second harmonic intensity 
	normalized by the laser energy as a function of square of applied Laplacian 
	electric field strength (symbols) together with the linear fit (solid curve).}
	\label{fig_EFISH_calibration}	
\end{figure}

To obtain the absolute value of the electric field a calibration is necessary. 
Signal intensities  at known electric field values are usually measured to 
determine the proportionality coefficient between $ I^{(2\omega)}/I^2_{laser} $ and $ E^2 $. 
Without the plasma the electric field is Laplacian and depends only on 
the discharge cell geometry and the voltage applied to the electrodes. 
Taking the discharge cell configuration into account, see figure~\ref{fig:cs_jet},
it is assumed that the electric field is uniform in the vicinity of the measurement points  
and corresponds to the capacitor like configuration with $ E=U/d $, 
where $ U $ is the applied voltage and $ d=1 $~mm is the inter-electrode gap. 
It was shown  \cite{lepikhin_electric_2020} that if a DC voltage is used to provide the 
Laplacian field the  slope of the calibration curve, $ I^{(2\omega)}/I^2_{laser} $ vs $ E^2 $, 
may be too low leading to an overestimation of the electric field  
in the discharge. It was proposed that the applied field is shielded by the charges 
generated due to tightly focused laser emission, thus the field in not Laplacian 
even without the discharge. If a DC voltage is applied, these charges are separated 
between the electrodes by the constant electric field, 
drift to the electrodes and, thus, do not recombine  forming an electric field, 
which shields the applied one. To avoid this problem it was proposed \cite{lepikhin_electric_2020} 
to use nanosecond  pulses at a low repetition rate instead of DC voltage. 
The same approach is used in this work. The calibration curve obtained with 
150~ns pulses with sub-breakdown amplitude is presented in figure~\ref{fig_EFISH_calibration}.
It is clearly seen that the PMT signal normalized by the laser energy linearly 
depends on the square of the applied electric field in accordance with \eref{eq_I_EFISH}.  
The parameters of the fit are used to obtain the longitudinal electric field value 
from the measured signal intensity  during the discharge. It should be noted here,
that the gas temperature - and therefore the particle density - is different in 
the discharge (see section \ref{sec:results}) than in the calibration performed at room temperature, this has to be taken into account.

\section{Experimental setup}

\begin{figure}
	\centering
	\includegraphics[width=\linewidth]{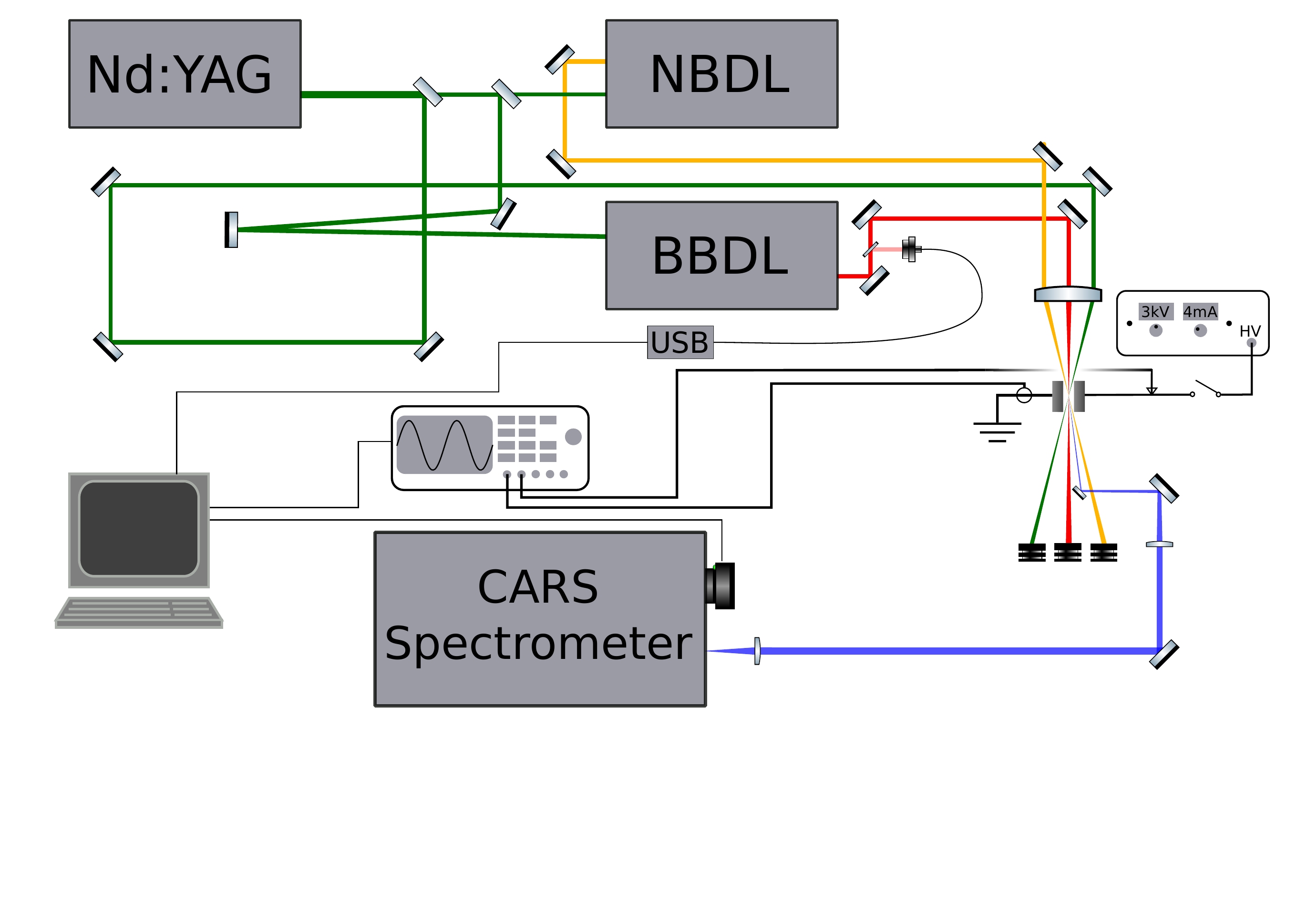}
	\caption{Optical setup of the CARS measurements. The broadband dye laser 
	(BBDL) and the narrowband dye laser (NBDL) are both pumped by the injection 
	seeded Nd:YAG laser. The spectrum of the BBDL is recorded during the measurements
	with a compact USB spectrometer (USB). Further information about the setup are given in the
	text.}
	\label{fig:optical_setup}
\end{figure}

\begin{figure*}
	\centering
	\includegraphics[width=\linewidth]{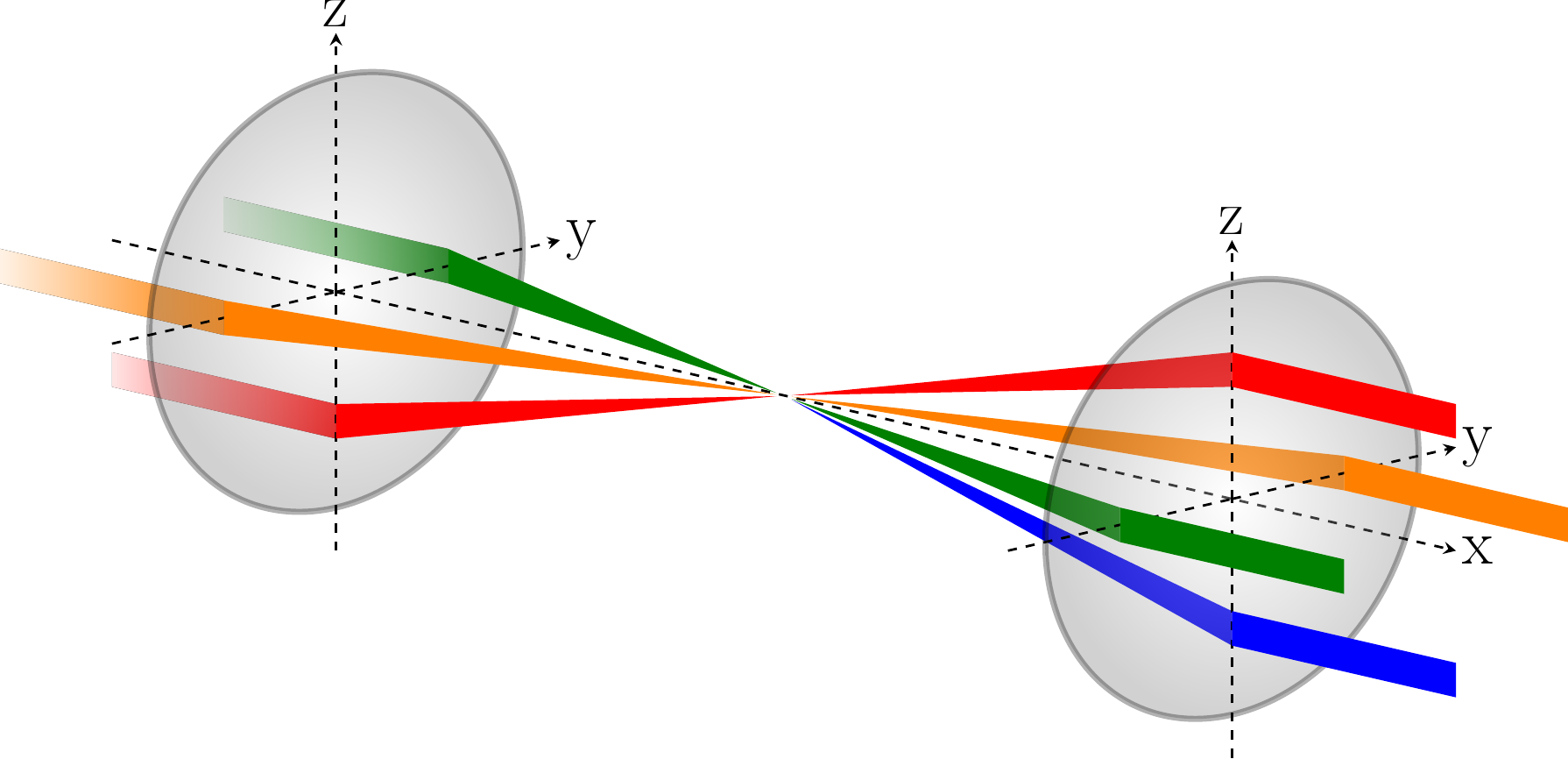}
	\caption{Three-dimensional phase matching in folded BOXCARS geometry. 
	For technical reasons in this work only the blue anti-Stokes beam is collimated (see also figure 
	\ref{fig:optical_setup}).}
	\label{fig:foldedBOXCARS}
\end{figure*}

\begin{figure}
		\centering
		\includegraphics[width=\linewidth]{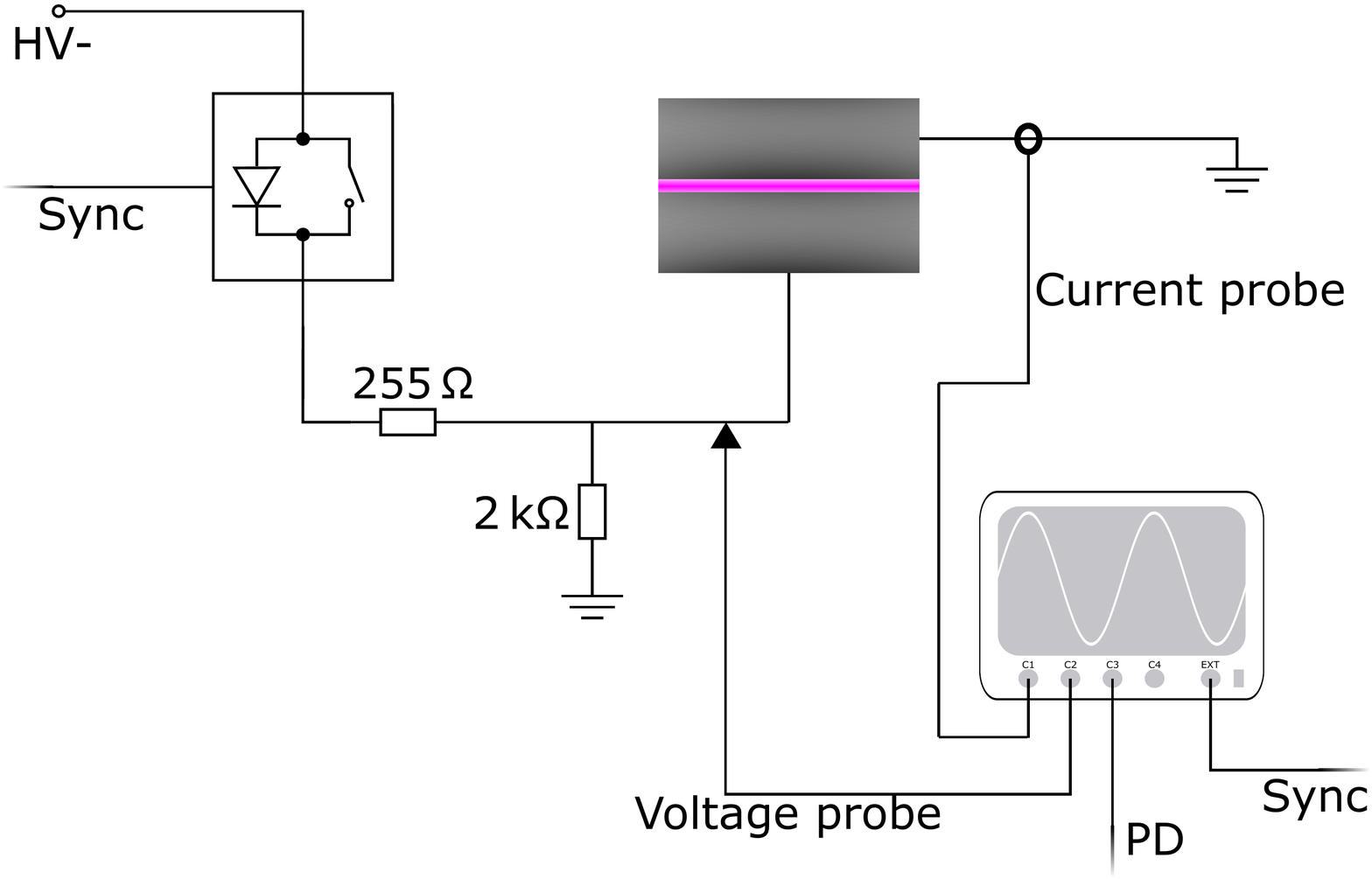}
		\caption{Electrical setup of the plasma jet.}
		\label{fig:electrical_setup}
\end{figure}

The experimental setup used in this work consists of optical setups for 
the CARS and E-FISH measurements and the electrical setup of the plasma jet.
The optical setups are synchronized with the discharge by a synchronization unit. \\

\subsection{Optical setup for the CARS measurements}
The optical setup - a modified version of the one used in
\cite{muller_ignition_2011} and \cite{bohm_determination_2016} for CARS-based electric field measurements - 
is shown in figure \ref{fig:optical_setup}. 
The second harmonic of an injection seeded Nd:YAG laser (\SI{532}{nm}) is used both as pump laser for two dye lasers and
as pump beam in the CARS process. One of the dye lasers is a narrowband dye laser
(NBDL), which is used as probe beam. 
The laser emission at \SI{560}{nm} is produced by a Rhodamine 6G dye solution and a 
spectral width of about \SI{0.05}{\per\centi\metre} is reached with a double grating 
configuration. Here, light with \SI{560}{nm} was chosen for the probe beam to be able
to perform CARS measurements in carbon dioxide and nitrogen simultaneously \cite{roy_dual-pump_2004} at a 
later stage.
In the broadband dye laser (BBDL) the double grating configuration for 
the wavelength selection is bypassed by a mirror to allow broadband operation.
The desired spectrum of the BBDL - which is monitored by a broadband spectrometer -
is reached by using a mixture of Pyrromethene 597
and Rhodamine B/101. Initially a mixture of Pyrromethene 597 and Pyrromethene 650
was used as in \cite{tedder_characteristics_2011}, but was found to drift to lower wavelengths
too quickly. As an attempt to increase the lifetime of the mixture Pyrromethene 650
was replaced by the two Rhodamine dyes. 
To adjust the temporal overlap of the three laser beams two delay stages are used:
One for the part of the Nd:YAG beam which is used in the CARS process and the
other one for the pump beam of the BBDL.
All three laser beams are first attenuated to pulse energies 
around \SI{5}{mJ} and then focused to the
probe volume by an achromatic lens with \SI{50}{cm} focal length. 
A three-dimensional folded BOXCARS \cite{eckbreth_laser_1996} geometry is used as depicted in figure 
\ref{fig:foldedBOXCARS} with a spatial resolution of about \SI{1}{cm} along 
the laser path and a few \SI{100}{\micro\metre} perpendicular - 
defined by the overlapping volume of the three laser beams.
The anti-Stokes signal beam passes through a low pass filter to reduce stray light from the lasers and is guided to the CARS spectrometer.
Furthermore, a photo diode is used to measure the timestamp of the laser on the oscilloscope, 
which also measures the current-voltage waveforms of the plasma jet.
The trigger of the Nd:YAG laser, the spectrometers and the oscilloscope (Lecroy WaveSurfer 510) are synchronized by a synchronization unit 
to allow the accumulation of single shot measurements.
Finally, all measured data are collected by a custom made DAQ software on the computer.

\subsection{Optical setup for the field measurements}
The emission of a  Nd:YAG  laser (EKSPLA SL234) at 1064~nm with a pulse duration 
of 100~ps is focused to the middle of the discharge gap by a lens  with focal 
length of 200~mm. The light at the second harmonic (532~nm) generated in presence 
of the external field is separated from the fundamental harmonic by a dichroic mirror 
and   detected by a photomultiplier tube (PMT; Hamamatsu H11901-210) with 
laser line filter (Thorlabs FL532-1) installed at its entrance. 
For the schematic view and more detailed description of the experimental setup 
used for the measurements of the electric field, please see \cite{lepikhin_electric_2020}, 
where a discharge with the same geometry was investigated. 

\subsection{Electrical setup and discharge jet}

\begin{figure}
	\centering
	\includegraphics[width=0.6\linewidth]{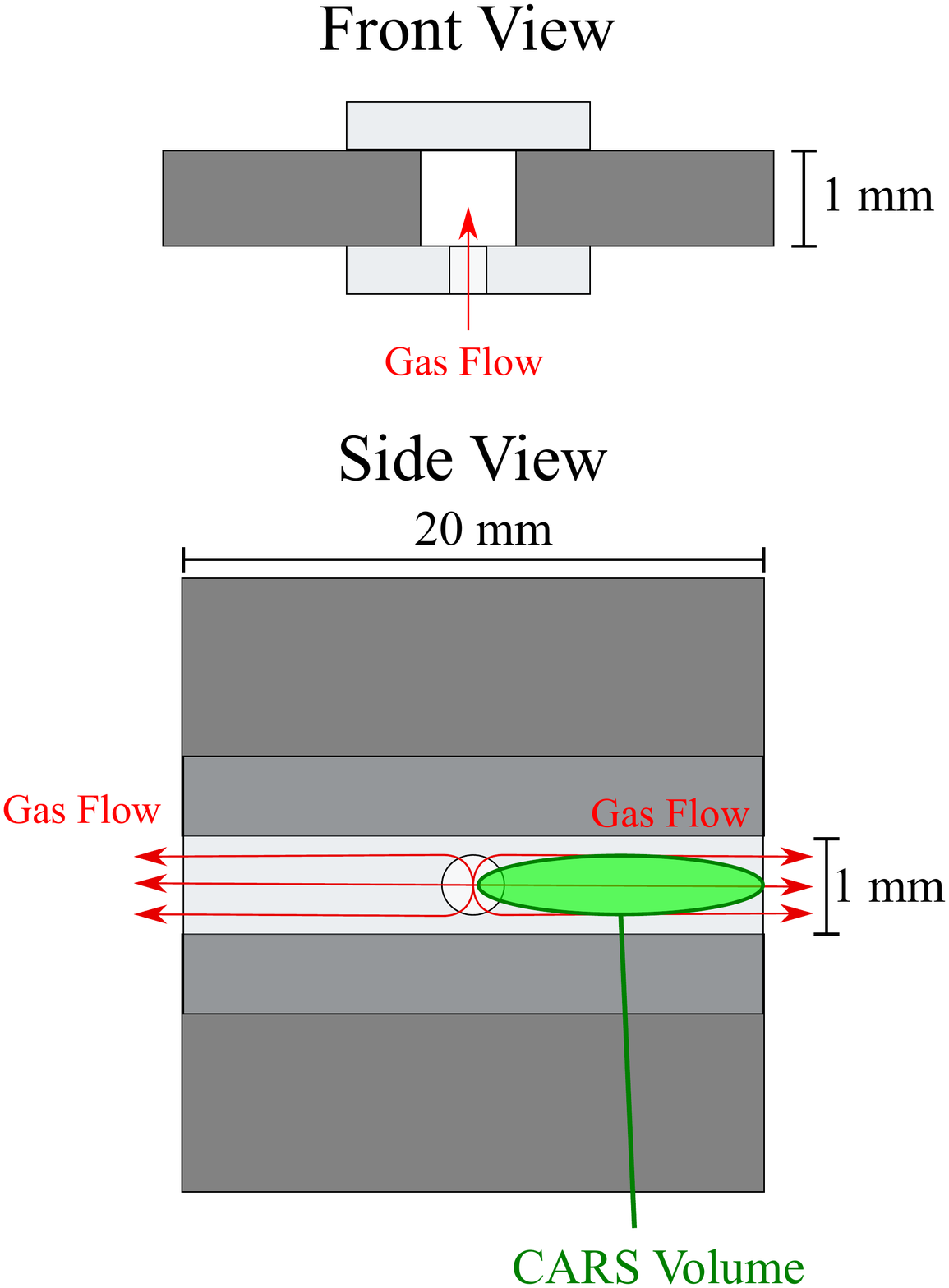}
	\caption{Front and side view of the discharge jet. The \SI{1}{mm} gap 
	between the electrodes is enclosed by two glass plates. One of the 
	glass plates has a small borehole in the middle which serves as inlet for 
	the gas flow. From the center of the discharge gap the gas travels to
	the exits on both sides yielding two identical discharge channels with 
	a cross section of \SI{1}{\square\milli\metre} and a length of \SI{10}{mm}.}
	\label{fig:cs_jet}
\end{figure}
The electrical setup of the plasma jet is depicted in figure \ref{fig:electrical_setup}. \\
A negative DC high voltage (Heinzinger LNC 6000-10neg) is applied to the input of a high voltage switch (Behlke HTS 81).
In order to protect the switch and increase the discharge stability the current is limited to $<\SI{30}{A}$ by a \SI{255}{\ohm} series resistor between switch and 
anode of the plasma jet. 
Additionally, a \SI{2}{\kilo\ohm} resistor ensures that residual charges are removed after each discharge pulse.
The voltage at the cathode relative to ground is measured by a high voltage probe (Lecroy PPE6kV)
and the current is measured between the anode and ground by a current probe (American Laser Systems, Model 711 Standard).
The Behlke switch is triggered by a \SI{1}{\kilo\hertz} TTL signal which is synchronized to the \SI{20}{\hertz}
trigger signal of the Nd:YAG laser through a frequency divider. 
The duration of the high voltage pulse can be controlled by the duration of 
the trigger pulses at a digital delay generator (DDG) with a minimum pulse length of \SI{150}{\nano\second}. \\
The plasma jet itself consists of two plan parallel molybdenum electrodes with an area of $\SI{1}{mm} \times \SI{20}{mm}$
and a distance of \SI{1}{mm}, which are positioned between two glass plates. 
To allow operation in pure nitrogen at pressures around \SI{200}{mbar} the electrode surfaces 
must be prepared carefully in order to avoid arcing. For this purpose the electrodes 
here were polished with a final grit size of P2000.\\
To illustrate the gas flow the front and the side view of the jet are shown in figure \ref{fig:cs_jet}. 
The gas enters the electrode gap through a hole in the lower glass plate 
and exits on both sides after being exposed to multiple discharge pulses due to the gas residence time. 
This design allows the laser beams to pass through the jet along the electrodes. 
It should be noted, that the measurement volume of the CARS setup is not 
in the center of the jet as there mostly unexcited gas would be measured, which was not 
exposed to the discharge yet. Instead, it stretches 
from the gas inlet in the middle to one of the exits.\\
To ensure a controlled atmosphere the jet is enclosed in a discharge chamber 
which allows controlling the pressure and the gas flow through the jet.

\section{Results and discussion}
\label{sec:results}
\begin{figure}
	\centering
	\includegraphics[width=\linewidth]{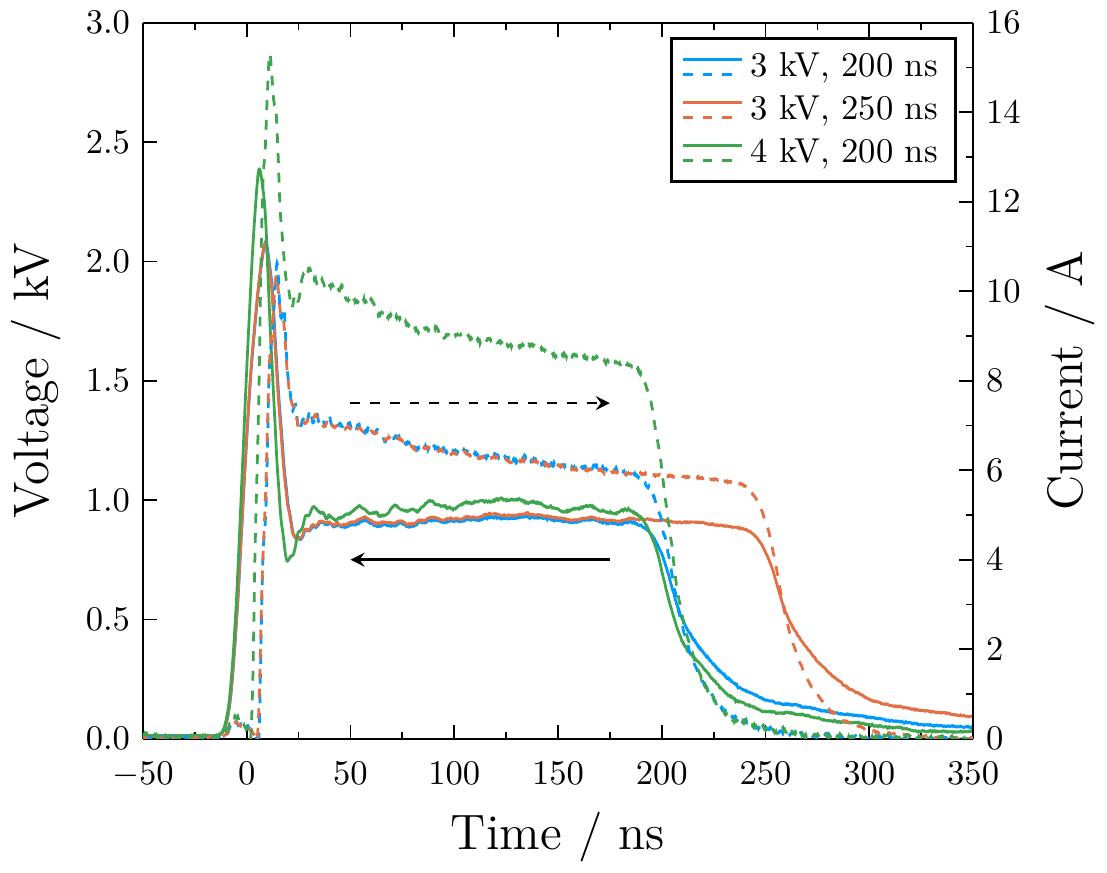}
	\caption{Voltage (solid lines) at the powered electrode and current (dashed lines) 
	to ground during the discharge for the different applied voltage pulses. While the 
	maximum peak voltage at the time of the breakdown increases for the higher applied 
	voltage it can be noted that the voltage drop over the discharge gap during the 
	following plateau is nearly the same for all three conditions. The current on the 
	other hand is significantly higher for the \SI{4}{kV} case. }
	\label{fig:VI_waveform}
\end{figure}

Three time resolved measurements are performed in pure nitrogen discharges for different voltage pulse forms.
The other input parameters are fixed for all three measurements:
The nitrogen flow is \SI{20}{sccm} at a pressure of \SI{200}{\milli\bar} and the pulse repetition frequency is \SI{1}{kHz}.
The pulse waveforms are illustrated in figure \ref{fig:VI_waveform} together with the resulting current waveforms. \\
In the following the different measurements will be referred to by the voltage applied by the DC high voltage generator (which is not the 
same as the measured amplitude voltages at the electrodes) and the pulse duration. \\ 
First the results during the discharge pulse will be discussed and afterwards the dynamics during the afterglow 
will be presented.

\subsection{Discharge phase}
\label{sec:discharge}
\begin{figure}[h!]
	\centering{\includegraphics[width=\linewidth]{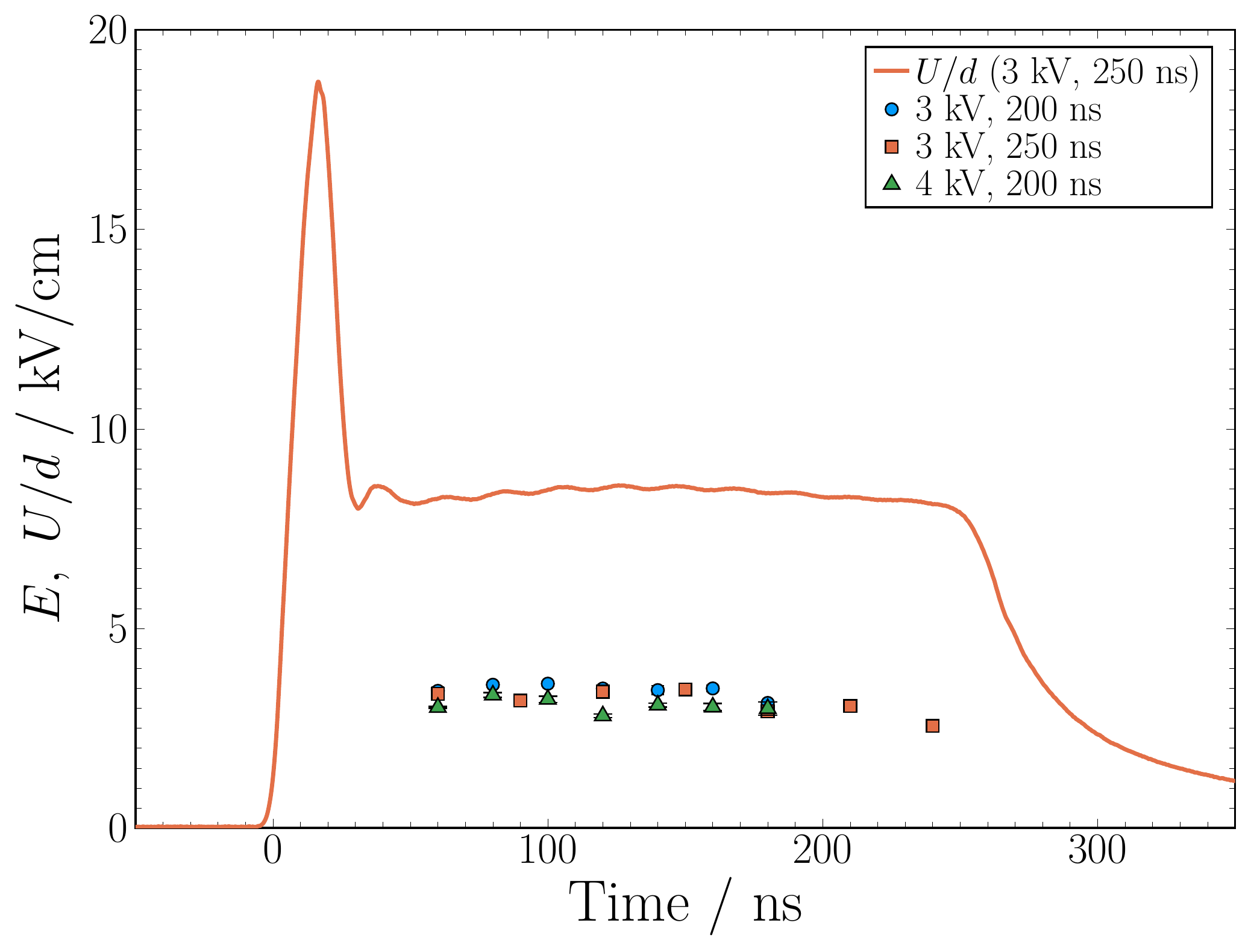}}
	\caption{Longitudinal electric field temporal development at 500~$\mu$m from 
	the cathode measured by E-FISH technique (symbols) for different HV pulses 
	compared with the electric field profile calculated as voltage over gap ratio for the 
	\SI{3}{kV}, \SI{250}{ns} measurement (line). As the voltage amplitude during the plateau phase
	is quite similar for all three measurements (see figure \ref{fig:VI_waveform}) only 
	one waveform is shown.
}
	\label{fig_EFISH_fields}	
\end{figure}

\begin{figure}[h]
	\centering
	\includegraphics[width=\linewidth]{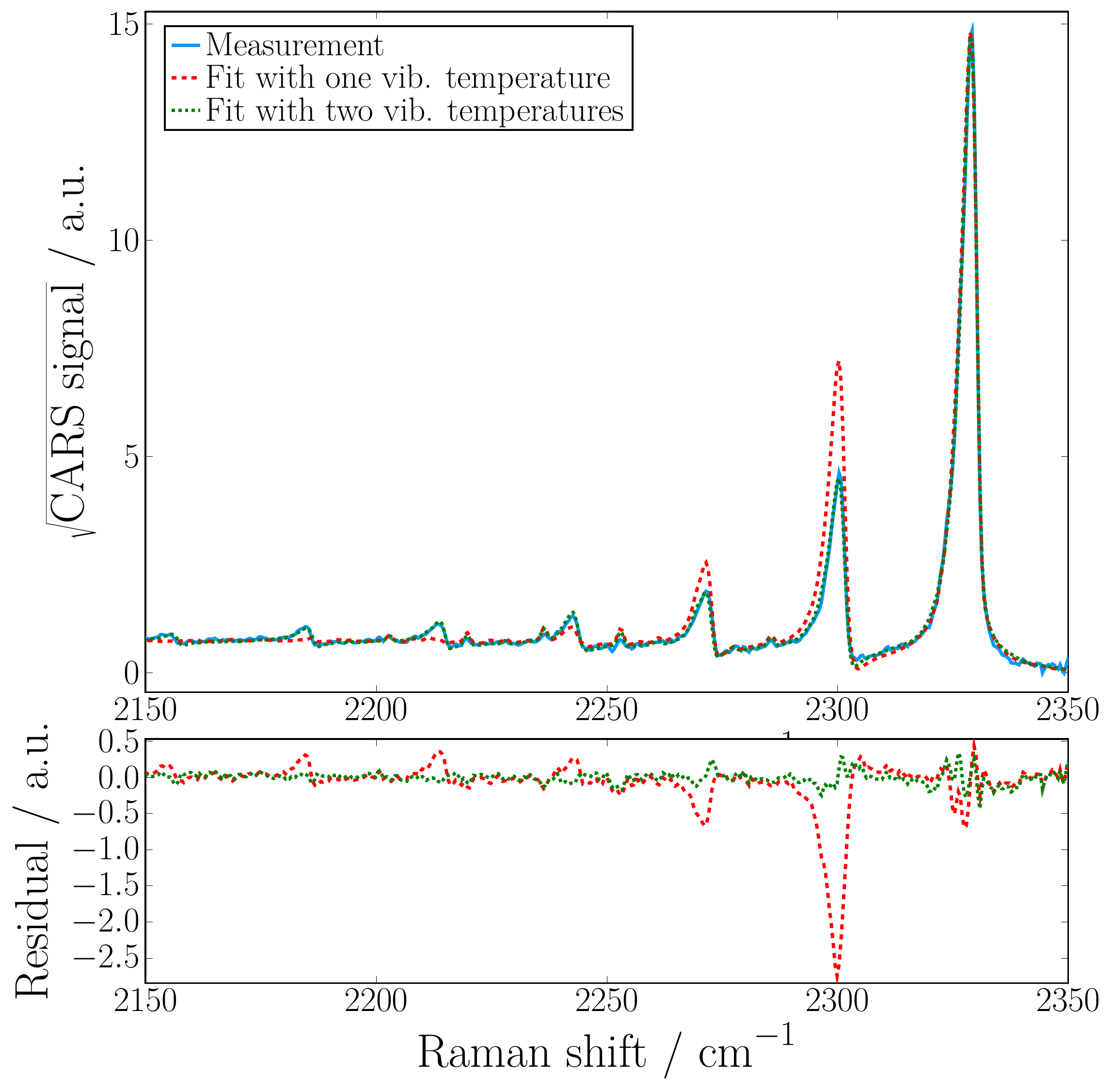}
	\caption{Comparison of a measured spectrum with theoretical spectra for Boltzmann distributed vibrational 
	states with $T_{vib}=\SI{1800}{\kelvin}$ and for a two-temperature distribution
	with $T_{vib,c}\approx\SI{1200}{\kelvin}$ and $T_{vib,h}\approx\SI{5200}{\kelvin}$.
	The spectrum corresponds to $t=\SI{300}{ns}$ in the measurement set for 
	an applied voltage of \SI{4}{kV} (approx. \SI{50}{ns} after the pulse).}
	\label{fig:TTDF_boltz_comp}
\end{figure}
\begin{figure}[h]
	\centering
	\includegraphics[width=\linewidth]{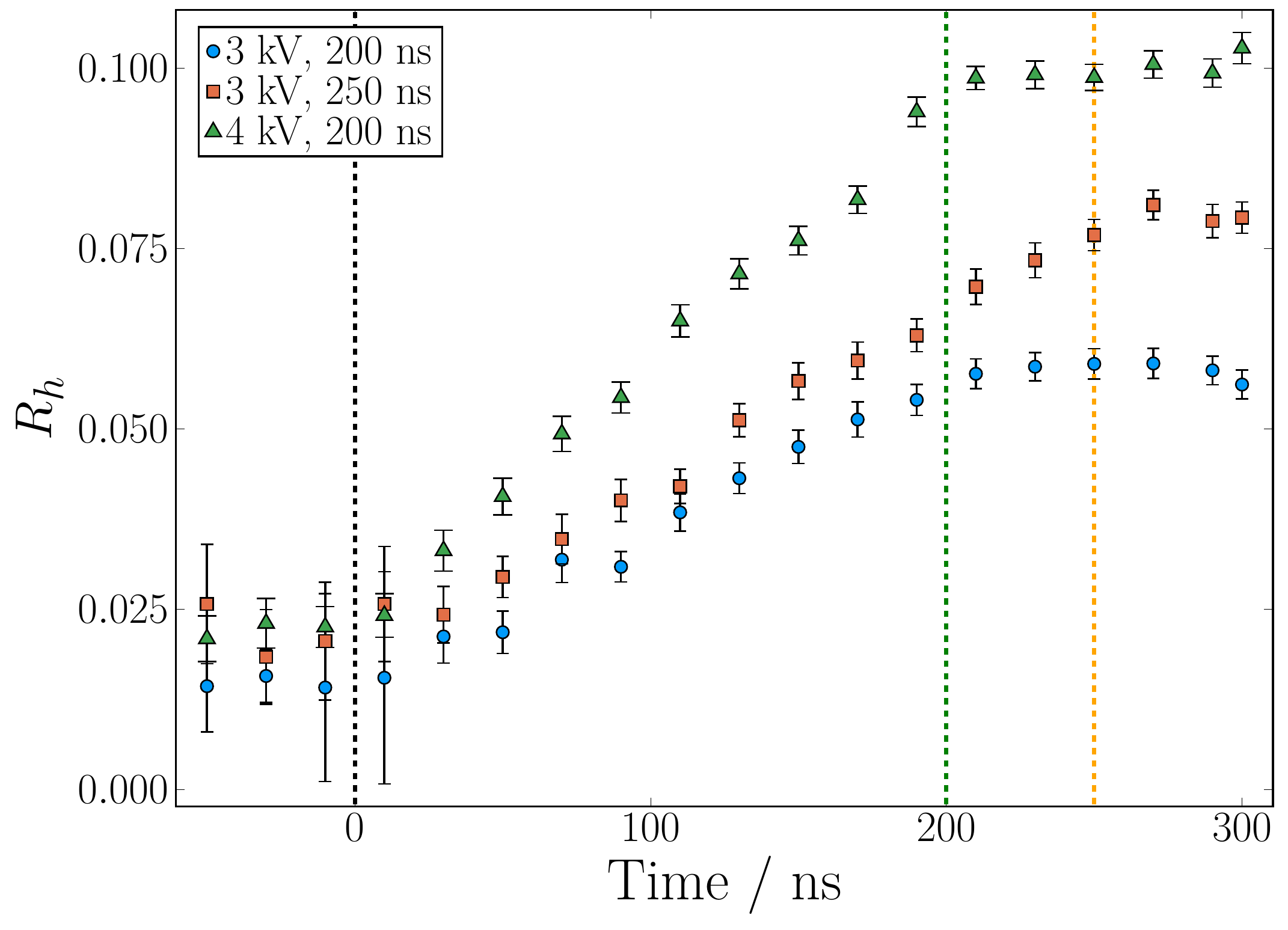}
	\caption{Time development of the fraction of vibrationally hot molecules 
	$R_h$ in \eref{eq:N_twotemp}. As guide for the eye the starting 
	point of the voltage pulses is marked by the black dashed line and the end points 
	of the \SI{200}{ns} and \SI{250}{ns} pulses are marked with the green and 
	the orange dashed lines respectively. The uncertainties of $R_h$ are determined 
	from the local Jacobian of the best fit.\label{fig:Rh_discharge}}
\end{figure}
\begin{figure}[h]
	\centering
	\includegraphics[width=\linewidth]{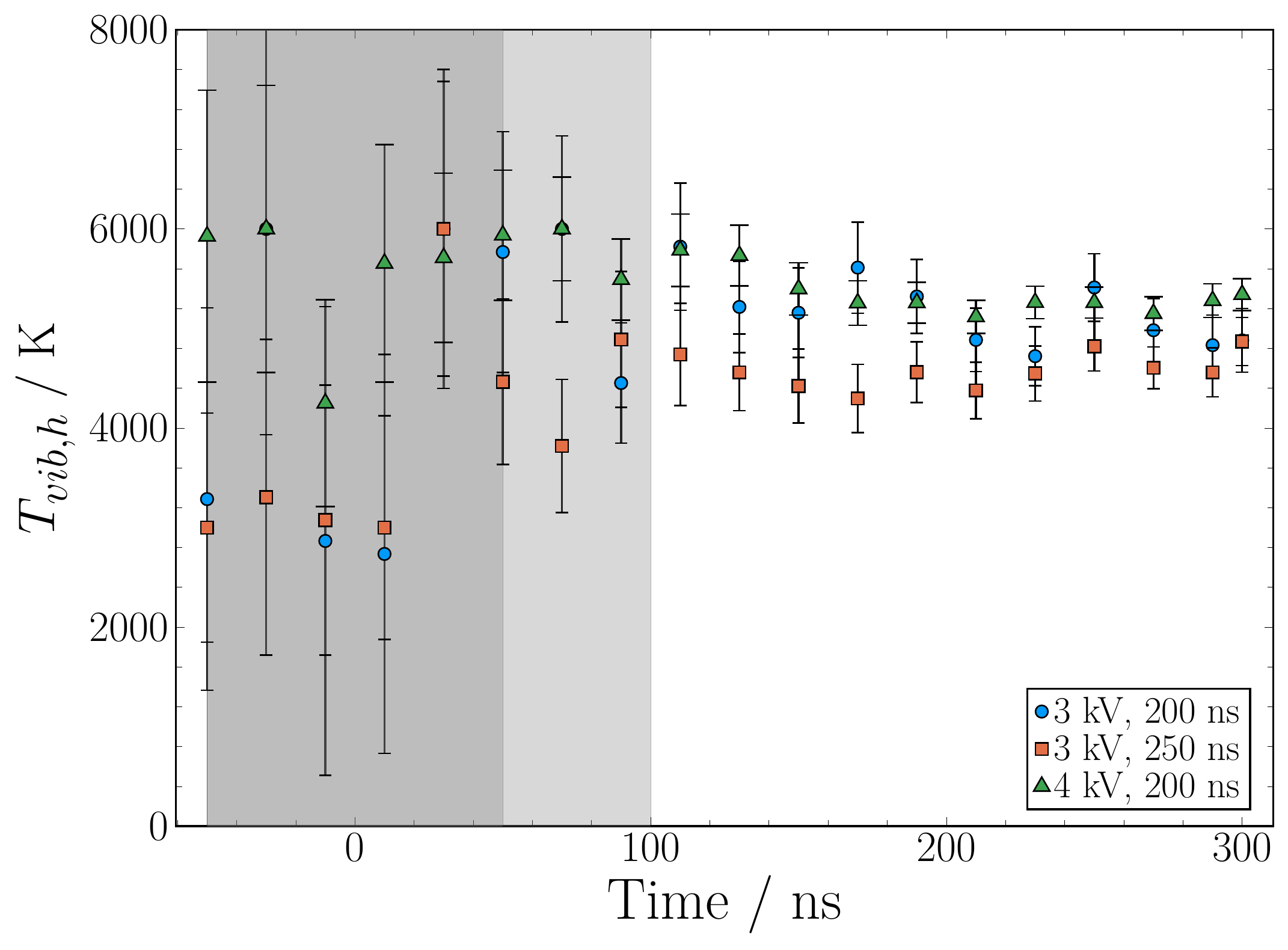}
	\caption{Time development of the temperature of the vibrationally hot 
	molecules $T_{vib,h}$ in \eref{eq:N_twotemp}. The gray areas are
	used to emphasize that temperatures for times lower than \SI{50}{ns} (\SI{4}{kV})
	or \SI{100}{ns} (\SI{3}{kV}) are not
	reliable for physical interpretation (see text). The uncertainties of $T_{vib,h}$
	are determined from the local Jacobian of the best fit.\label{fig:Tvh_discharge}}
\end{figure}
The electric field during the current plateau is shown in figure~\ref{fig_EFISH_fields} 
for  different DC voltages used to feed the high voltage switch and different 
HV pulse duration. The electric field has constant value of  $ \approx 3.5 $~kV/cm 
or $ \approx 81 $~Td during the plateau, which is about the same for different pulse duration. 
Moreover, the electric field amplitude only very weakly depends on the DC voltage as well. 
Despite the different DC voltage, after the breakdown the 
cathode voltage is about the same, see figure~\ref{fig:VI_waveform}. 
This can be explained by a higher electron density generated during the breakdown 
at the higher DC voltage leading to a higher electric current and consequently a higher 
voltage drop at the series resistor. According to calculations with Bolsig+ \cite{hagelaar_solving_2005}
using the IST-Lisbon data set \cite{alves_ist-lisbon_2014} the measured field corresponds to a mean electron energy of 
\SI{1.8}{eV} or an electron temperature of $T_e=\SI{1.2}{eV}$. \\
To demonstrate the necessity of the TTDF \eref{eq:N_twotemp} in figure \ref{fig:TTDF_boltz_comp}
a measured spectrum with its corresponding TTDF fit is compared with a theoretical spectrum 
for a single vibrational temperature. From the figure it is obvious, that 
a single vibrational temperature will not be able to reproduce the measured spectra
during or shortly after the discharge.
The time resolved  CARS fitting results for a two temperature distribution according to  
\eref{eq:N_twotemp} are shown in figures \ref{fig:Rh_discharge} and \ref{fig:Tvh_discharge}.
The uncertainties of the fitting parameters $R_h$ and $T_{vib,h}$ are determined from the 
local Jacobian of the corresponding best fit solutions.
During the discharge pulse the fraction of vibrationally hot molecules, $R_h$, increases 
linearly. This agrees well with the essentially constant current and electric field 
after the ignition suggesting constant excitation conditions during the whole discharge pulse.
It is clearly visible, that the slope of $R_h$ is the highest for the pulse with the
higher current amplitude (\SI{4}{kV} applied DC voltage) while the two 
measurements with \SI{3}{kV} applied voltage share a lower slope and differ only
by a small offset due to different initial conditions.
As the electric field is the same in all three cases, the difference in 
slope should be mainly caused by a higher electron density in the \SI{4}{kV} case.
A more detailed discussion of the slope, $\dot{R}_h$, is given in the companion paper \cite{kuhfeld_vibrational_nodate}.
The fact, that one of the measurements is performed with a longer pulse duration,
is also reflected in figure \ref{fig:Rh_discharge} by a longer rise time of $R_h$ 
(note the two vertical lines at about \SI{200}{ns} and \SI{250}{ns} marking 
the corresponding ends of the pulses). \\
In regard to the temperature of the hot population it has to be noted that the values
at the beginning, when the amount of excited molecules is small, are not reliable 
or suitable for a direct interpretation. Firstly, at these time only states up to
$v=3$ can be detected as the densities of the other states are below the detection 
limit. Secondly, \eref{eq:N_twotemp} is proposed to analyze the characteristics
of the newly excited molecules (for further details the reader is referred to 
the companion paper \cite{kuhfeld_vibrational_nodate}). Before the discharge pulse there are no "newly" excited
molecules yet and the non-zero value of the fitting parameter $R_h$ only accounts
for the small deviation from a normal Boltzmann distribution remaining from 
the previous voltage pulse still visible in the 
excited states $v=1-3$ and $T_{vib,h}$ suffers a large uncertainty, as it strongly 
depends on the higher states, which are not detected. 
So, to be able to use the two-temperature distribution in its original intent 
the newly excited molecules need to overtake the remaining deviation from a 
Boltzmann distribution. The criterion used here and in \cite{kuhfeld_vibrational_nodate} is 
that $R_h$ needs to at least surpass two times its initial value from before the discharge, 
so that the majority of the molecules belonging to the hot distribution consists 
of molecules excited during the current discharge pulse. The data points which 
are excluded from further discussion by the mentioned criterion are marked by 
a gray background in figure \ref{fig:Tvh_discharge}. Please note, that this criterion
agrees quite well with the uncertainties of $T_{vib,h}$. 
Considering the previous remarks it stands out that $T_{vib,h}$ stays constant for all three measurements
in the usable range. While the \SI{200}{ns}-pulse measurements for \SI{3}{kV} with \SI{5200\pm 300}{\kelvin}
and the one for \SI{4}{kV} with \SI{5300 \pm 200}{\kelvin} are very close to each 
other, the measurement for the \SI{250}{ns} pulse deviates with \SI{4600 \pm 200}{\kelvin}.
This deviation is probably caused by a worse signal-to-noise ratio during that 
particular measurement set. The dye mixture of the broadband Stokes laser needs to be 
refreshed between measurement sets and while the spectrum of the Stokes beam
is included in the calculation of the CARS spectra, the fluctuations of the 
Stokes beam result in different signal-to-noise ratios in the measured CARS signals
for the different measurement sets. As this mainly affects the higher vibrational 
states due to their weaker signal, the effect is stronger on the fitting parameter
influenced by those states, namely the hot vibrational temperature $T_{vib,h}$. 
$R_h$ on the other hand corresponds to the total amount of newly excited molecules
and as the higher vibrational states only contribute by a small number - even at 
temperatures around \SI{5000}{K} - the effect is small. This explains why there 
is no deviation from the expected behavior visible in figure \ref{fig:Rh_discharge}.
As one would expect the rotational temperature stays constant on time 
scales of the discharge with about $T_{rot}=\SI{330 \pm 8}{\kelvin}$ (mean value
and standard deviation over all three measurement sets). Likewise, the vibrational
temperatures of the cold bulk molecules stays nearly constant during the discharge 
pulse as well, but differ slightly for the different measurement sets. The two 
measurements with \SI{3}{kV} applied DC voltage are quite close to each other with
\SI{1190 \pm 40}{\kelvin} (\SI{200}{ns} pulse) and \SI{1180 \pm 40}{\kelvin}
(\SI{250}{ns} pulse). 
For the \SI{4}{kV} case the vibrational background temperature is slightly higher
with \SI{1300 \pm 50}{\kelvin}.

\begin{figure}
	\centering
	\includegraphics[width=\linewidth]{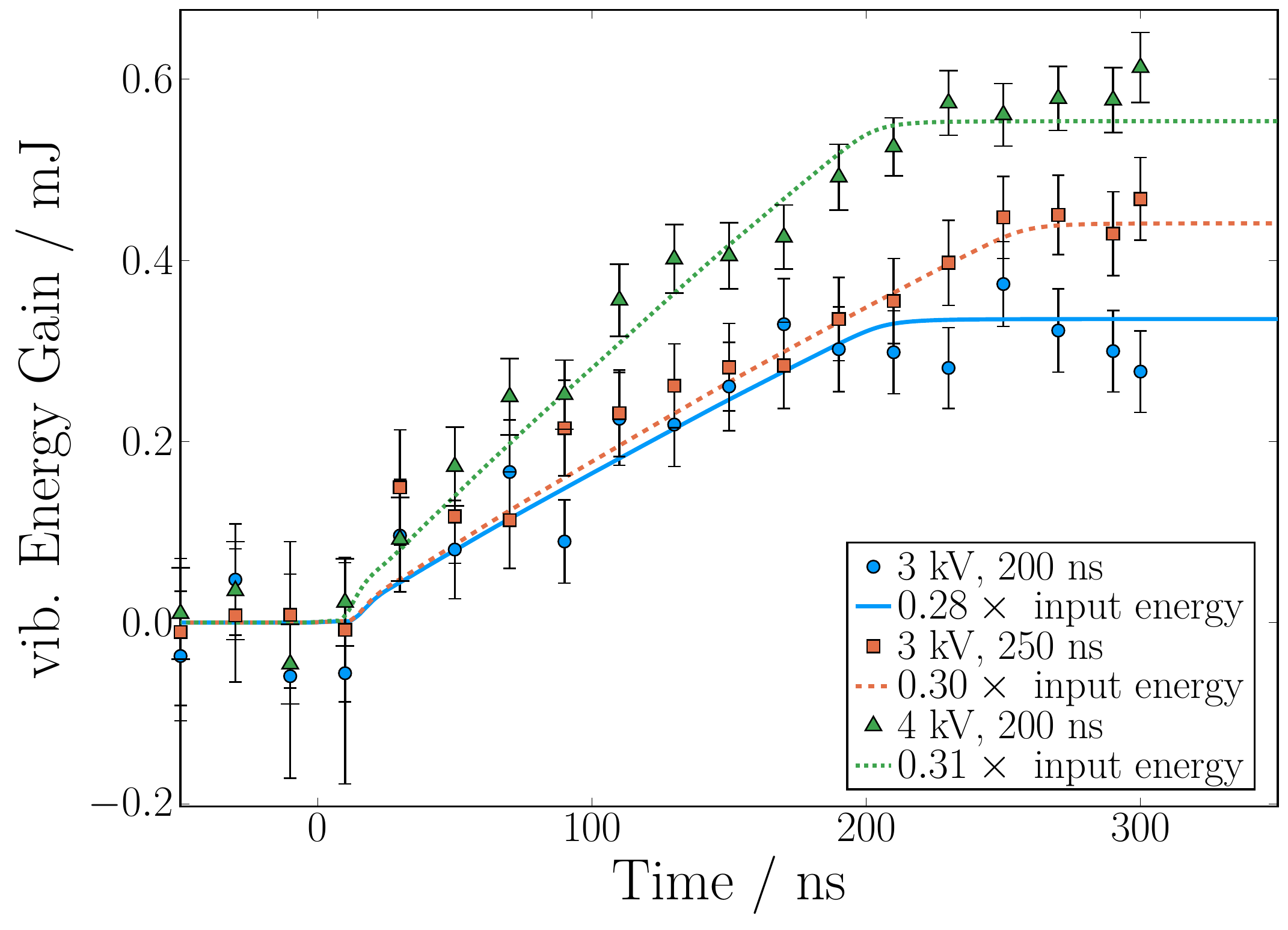}
	\caption{Gain of vibrational energy during the discharge pulse. 
	The average of the first three data points in each data set is used as baseline, 
	i.e. the initial vibrational energy before the discharge. The solid lines refer to 
	the electrical energy coupled into the plasma obtained from the voltage and 
	current waveforms in figure \ref{fig:VI_waveform}  
	multiplied by the factors given in the legend.}
	\label{fig:vib_gain}
\end{figure}

As the densities of the vibrational excited states are known from the fits, the gain of 
vibrational energy in the plasma volume during one discharge pulse can be calculated relative to the average energy 
of the first three measurement points (see figure~\ref{fig:vib_gain}).
By a comparison with the energy coupled into the plasma obtained from 
the voltage and current waveforms in figures~\ref{fig:VI_waveform}
it can be seen, that the energy coupled into 
vibrational modes is proportional to the total energy at all times 
during the discharge pulse. The corresponding factor is found close to 0.3 in 
figure~\ref{fig:vib_gain} for all measurement sets, indicating that the coupling 
efficiency is roughly \SI{30}{\percent} for all investigated conditions. 
This value agrees well with the \SI{30}{\percent} reported by 
Montello \etal\cite{montello_picosecond_2013} and Deviatov \etal\cite{deviatov_investigation_1986}.

\subsection{Afterglow}
\label{sec:afterglow}

\begin{figure}
	\centering
	\includegraphics[width=\linewidth]{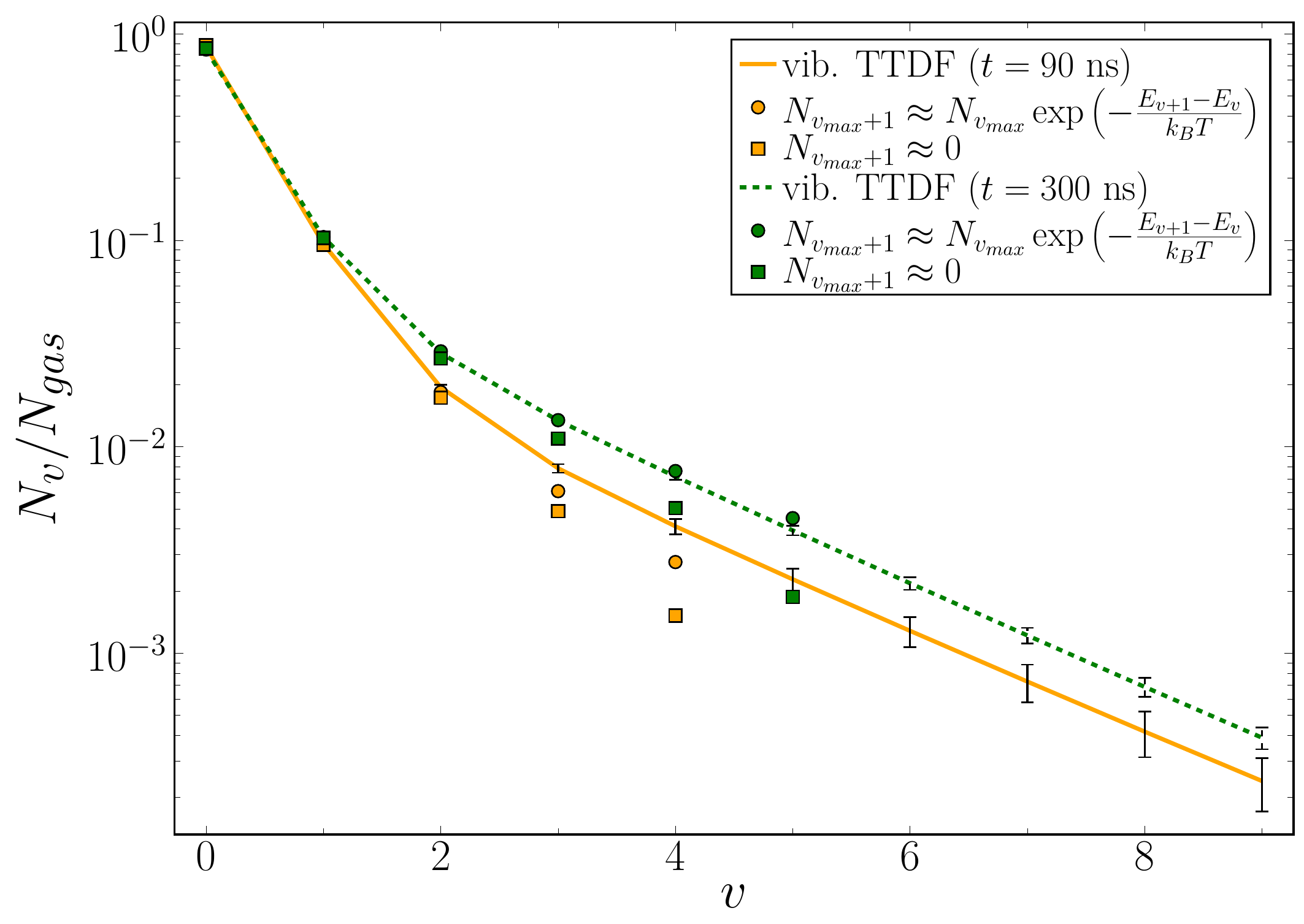}
	\caption{Comparison of the population densities obtained by the two-temperature
	distribution and the Q-branch-only fit with the two different evaluation 
	methods. The solid and dashed lines correspond to the TTDF at different points in 
	time, $t=\SI{90}{ns}$ and $t=\SI{290}{ns}$ respectively. The symbols in the 
	corresponding colors denote the results at the same times from the Q-branch-only 
	fits evaluated with the newly proposed method (circles) and the traditional
	one (squares).}
	\label{fig:approx_comparison}
\end{figure}

In the period between two pulses \eref{eq:N_twotemp} is not suitable to 
describe the VDF of the nitrogen molecules as interaction between the bulk 
and the excited molecules makes it difficult to distinguish effects on the 
fitting parameters $R_h$ and $T_{vib,h}$. Furthermore, it is not given, that 
the hot molecules - if this definition can still be applied - follow a distribution
which can be approximated by a Boltzmann distribution. For this reason the fitting
scheme discussed in the end of section \ref{sec:CARS_spectra} is used which yields the population
differences between two neighboring vibrational states (here normalized 
to the total particle density of nitrogen) $\frac{N_v - N_{v+1}}{N_{gas}}$  
as output parameters by considering only the Q branch transitions. 
The population differences $\Delta N_v = N_v - N_{v+1}$ - in the following $\Delta N_v$
is always assumed to be in arbitrary units - obtained by the fit could now be processed further 
by either making assumptions on the population difference between the ground and the
first excited state \cite{shaub_direct_1977} or by setting the population of the upper 
state for the highest detectable transition to zero \cite{montello_picosecond_2013}:
$\Delta N_{v_{max}} = N_{v_{max}} - N_{v_{max}+1}\approx N_{v_{max}}$.
With $v_{max}$ as starting point the population densities can then simply be obtained 
by $N_v = N_{v+1} + \Delta N_v$. Finally, they are normalized to $\sum_v N_v$ effectively
yielding the fraction of molecules in state $v$ compared to the gas density.
Obviously, the assumption $N_{v_{max}}\approx 0$ introduces a certain error into the 
analysis. To estimate this error one may take a look at the ratio of two neighboring states
following a Boltzmann distribution $N_{v+1}/N_v \approx \exp\left(-\frac{\hbar\omega_e}{k_B T_{vib}}\right)$
(neglecting the anharmonicity for simplicity). At room temperature this ratio is indeed much smaller than 
one and the above approximation is valid. But as it was shown in the previous section
that the highest vibrational states can have a temperature of about \SI{5000}{K} leading to
$N_{v+1}/N_v \approx 0.5$. Here the approximation is clearly not valid anymore. Therefore,
in this work $N_{v_{max}+1}$ is extrapolated from $N_{v_{max}}$ so that one obtains for the
population difference 
\begin{eqnarray}
	\label{eq:new_approx}
	\Delta N_{v_{max}} &\approx N_{v_{max}} -N_{v_{max}}	\exp\left( -\frac{E_{v_{max}+1}-E_{v_{max}}}{k_B \tilde T_{vib}}\right)  \\\nonumber
	&\approx N_{v_{max}} \left[1- \exp\left( -\frac{E_{v_{max}+1}-E_{v_{max}}}{k_B\tilde T_{vib}}\right)\right]
\end{eqnarray}
where $\tilde T_{vib}$ is estimated from the ratio of the two highest measured population differences:
\begin{equation}
	\tilde T_{vib} \approx - \frac{E_{v_{max}}-E_{v_{max}-1}}{k_B \ln\frac{\Delta N_{v_{max}}}{\Delta N_{v_{max}-1}}}
	\label{eq:Tvib_approx}
\end{equation}
In figure \ref{fig:approx_comparison} our approximation \eref{eq:new_approx} is compared with the traditional approximation
\mbox{$N_{v_{max}}\approx 0$} and the full TTDF fit from the previous section for two different points in time 
when the TTDF is assumed to be valid - during the discharge and shortly after. As can be seen quite 
clearly the agreement between the new method and the TTDF is significantly better than for 
the traditional approach. The small residual deviation at $t=\SI{90}{ns}$ is probably 
caused by the effect of the "bend" of the distribution function between the cold and the hot 
parts on the estimation of $\tilde T_{vib}$ in \eref{eq:Tvib_approx} - leading to a slight underestimation
of $\tilde T_{vib}$. It should be stressed, that while an underestimation of $N_{v_{max}}$ by a factor of 
about $2$ in the approximation $\Delta N_{v_{max}}\approx 0$ might be considered acceptable under some conditions,
this error amplifies when the data are used to extrapolate to higher vibrational states.
Therefore, \eref{eq:new_approx} and \eref{eq:Tvib_approx} are used in this work for the CARS spectra in the afterglow
where the shape of the distribution function is not known \textit{a priori}.
\begin{figure}
	\centering
	\includegraphics[width=\linewidth]{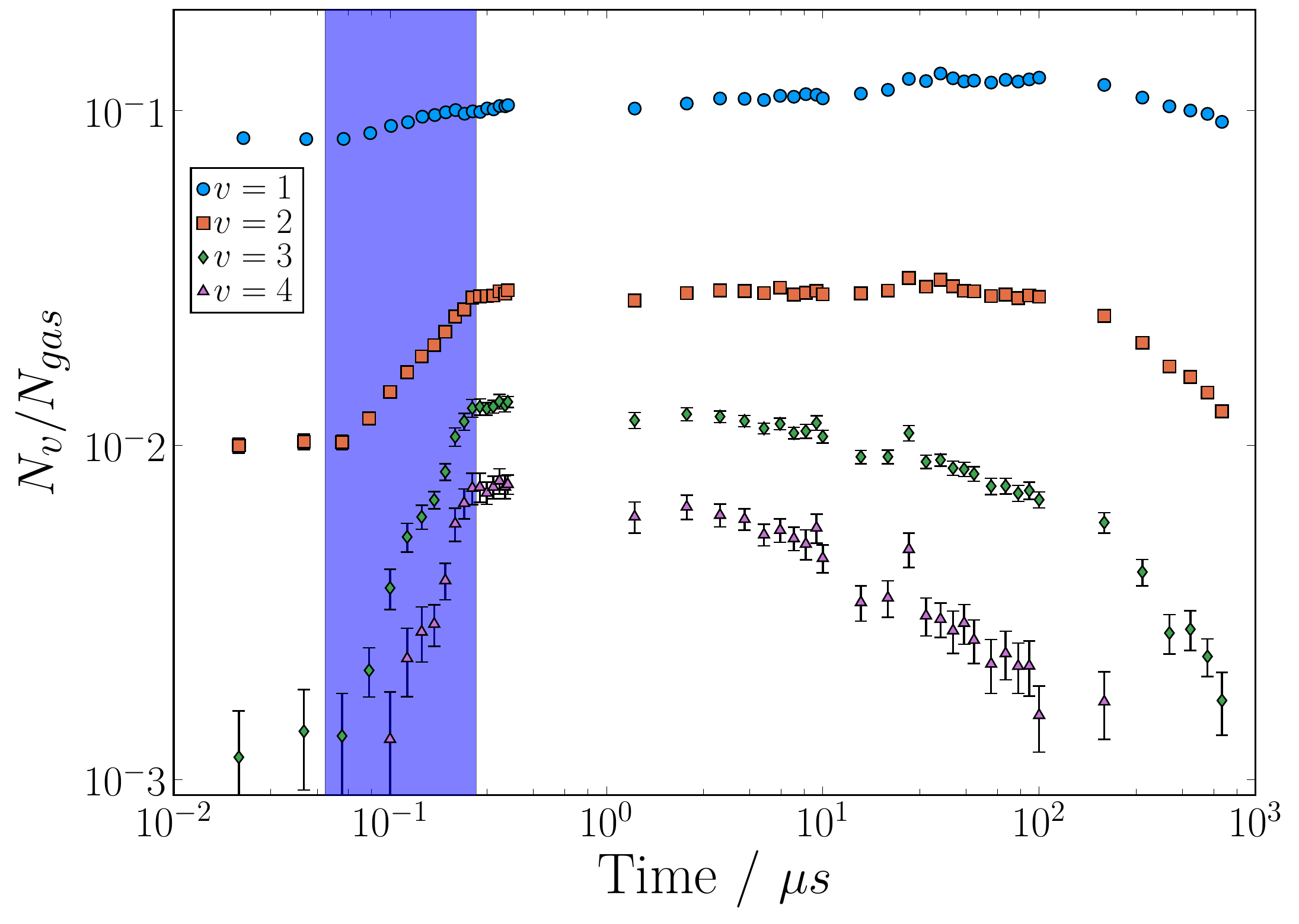}
	\caption{Relative population densities for $v=1,2,3,4$ during the discharge and
	in the afterglow for \SI{4}{kV} applied DC voltage and \SI{200}{ns} pulses. The blue
	area marks the time of the discharge pulse.}
	\label{fig:densities_afterglow}
\end{figure}%

\begin{figure}
	\centering
	\includegraphics[width=\linewidth]{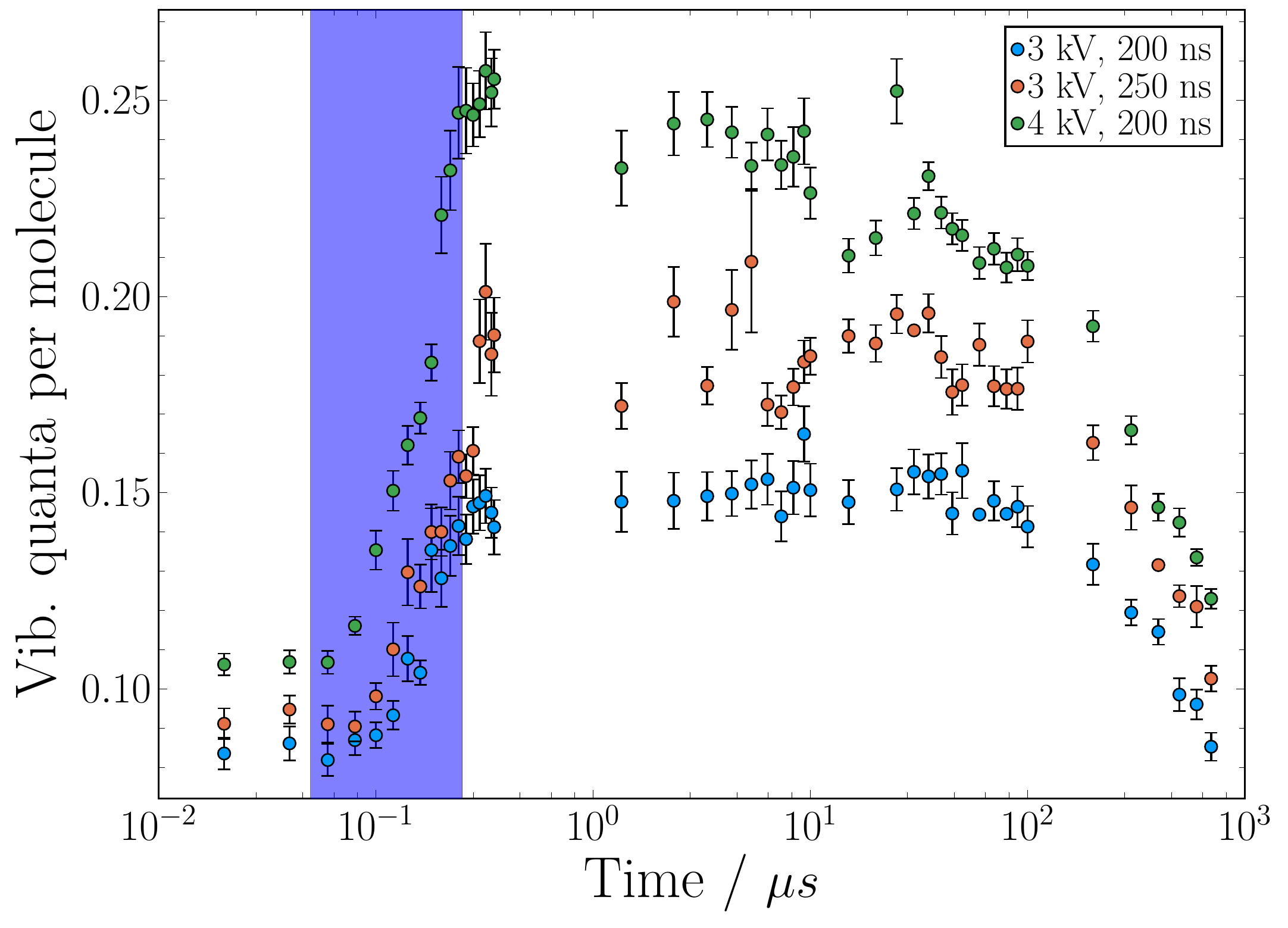}
	\caption{Vibrational quanta per molecule for all three measurements. The blue 
	rectangle marks the time during the discharge pulse.}
	\label{fig:vib_quanta}
\end{figure}

In figure \ref{fig:densities_afterglow} the relative population densities from the
states 1 up to 4 are shown for the measurement with \SI{4}{kV} applied DC voltage. 
While the population of the first vibrational state increases
and the one of the second stays approximately constant, the higher states decrease on
timescales smaller than about \SI{100}{\micro\second}. These dynamics can be attributed
to V-V transfer between the molecules \cite{kuhfeld_vibrational_nodate}. The general decrease of all excited states
on longer timescales was found to 
be mainly due to deactivation at the walls \cite{kuhfeld_vibrational_nodate}. As one can already guess
from figure \ref{fig:densities_afterglow} the number of vibrational quanta does not 
seem to increase in the afterglow due to deexcitation of electronically excited
molecules as it was observed by other groups in their discharges \cite{montello_picosecond_2013}.
This is shown explicitly for all three measurement sets in figure \ref{fig:vib_quanta}. 
The number of vibrational quanta per molecule stays approximately constant during the V-V transfer
phase and finally decreases to its original value due to losses at the walls with reasonable decay 
times \cite{kuhfeld_vibrational_nodate}.

\section{Conclusion}
In this work, a plasma is generated by applying high voltage pulses of \SI{200}{ns} 
and \SI{250}{ns} duration to plan-parallel molybdenum electrodes in nitrogen at \SI{200}{mbar}. 
In contrast to DBD discharges, more commonly used at these pressures, 
the metal electrodes allow for a nearly constant conduction current
after the ignition phase of the plasma until the end of the pulse. 
The electric field in these conduction current plateaus is measured by the 
E-FISH method and found to be constant during the discharge with a value around
\SI{81}{Td} for all conditions investigated here.
CARS spectra were measured for three different high voltage pulses. 
To extract information from those spectra more reliable than from an 
analysis of the separate peak intensities a fitting routine is developed, 
motivated by existing CARS fitting routines for combustion systems in thermal 
equilibrium \cite{luthe_algorithms_1986, farrow_comparison_1982,palmer_carsft_1989-1}
and benchmarked against the popular CARSFT Fortran77 code\cite{palmer_carsft_1989-1}. Our fitting procedure 
allows for arbitrary vibrational distribution functions to account for different 
conditions during the discharge phase and in the afterglow.
It was found, that the vibrational
excitation during the discharge phase can be described by a two-temperature 
distribution function consisting of a cold bulk with a near-equilibrium temperature
$ T_{vib,c} \approx \SI{1300}{\kelvin} > T_{rot}$ making up the majority of the molecules
and a hot part accounting for the molecules which are excited during the discharge 
pulse. The fraction of hot particles $R_h$ increases close to linearly with time and 
the hot temperature $T_{vib,h}$ stays constant. This indicates constant excitation 
conditions during the  pulses and agrees well with the constant electric field
obtained by the E-FISH measurements and the quasi constant 
currents after the ignition phase (see figure \ref{fig:VI_waveform}) both 
implying a constant electron density. The same behavior is seen in \cite{muller_ignition_2011} where
similar discharge conditions (conducting electrodes, atmospheric pressure and millimeter gap size) are
investigated.
The nearly constant plasma conditions allow an easy analysis of the vibrational
dynamics and a comparison with corresponding constants in the literature.
These are performed in a companion paper \cite{kuhfeld_vibrational_nodate} and the reader is 
referred to there for more details.
It can be summarized, that CARS is a complicated but powerful technique for measuring vibrational
distribution functions in plasmas. Especially, the possibility to measure multiple 
species at once with the same experimental setup will prove extremely useful for 
investigating the influence of vibrational excitation on chemical and catalytic 
processes. These prospects will be investigated in future works with the same setup
as introduced here. \\
Additionally, the application of CARS to an excited medium like a plasma itself provides 
access to further open research questions and opportunity for investigation. E.g. the long living
excited states in a plasma might influence the non-resonant susceptibility depending 
on the amount of excited species as it was observed for third harmonic generation 
in a pre-excited gas \cite{fedotov_third-harmonic_2000}. While the non-resonant part
of the susceptibility is ignored by many works it can be important for higher vibrational
states where the resonant and non-resonant parts are of comparable amplitude.

\section*{Acknowledgements}
This project is supported by the DFG (German Science Foundation) within the 
framework of the CRC (Collaborative Research Centre) 1316 "Transient atmospheric
plasmas - from plasmas to liquids to solids". \\
Furthermore, the authors would like to thank Igor Adamovich, Kraig Frederickson and Ilya Gulko for sharing
the CARSFT code and providing insights about the analysis of CARS spectra.

% With BibTex
\section*{References}
\bibliography{references}

\end{document}